\documentclass[
hf,
]{ceurart}
\usepackage[numbers]{natbib}

\usepackage{multirow}
\usepackage{hyperref}
\hypersetup{
    colorlinks=true,
    linkcolor=Crimson,
    filecolor=magenta,      
    urlcolor=teal,
    }
\usepackage[nameinlink, capitalize, noabbrev]{cleveref}

\usepackage{amsmath,amssymb,amsfonts}
\usepackage{xcolor}
\usepackage[linesnumbered, ruled]{algorithm2e}

\SetKwInput{KwInput}{Input}                
\usepackage{amsmath,amsfonts,amssymb,mathrsfs}

\begin{document}

\let\WriteBookmarks\relax
\def\floatpagepagefraction{1}
\def\textpagefraction{.001}
\shorttitle{ }
\shortauthors{ }
\title{A hybrid algorithm based on Community Detection and Multi-Attribute Decision-Making for Influence Maximization}

\conference{Published in \emph{Computers \& Industrial Engineering} 120 (2018): 234-250 -  \url{https://doi.org/10.1016/j.cie.2018.04.049}}
\author[1]{Masoud Jalayer}[%
orcid=0000-0001-8013-8613,
email=masoud.jalayer@polimi.it,
]


\author[2]{Morvarid Azheian}

\author[2]{Mehrdad Mohammad Ali Kermani}[%
email=kermani.m.a.m.a@gmail.com,
]

\address[1]{Department of Management, Economics and Industrial Engineering, Politecnico di Milano, Via Lambruschini 4/b, 20156, Milan, Italy}

\address[2]{Department of Progress Engineering, Iran University of Science and Technology, Narmak 16846, Tehran, Iran}


\maketitle
\begin{abstract}
Influence maximization problem is trying to identify a set of $K$ nodes by which the spread of influence, diseases or information is maximized. The optimization of influence by finding such a set is NP-hard problem and a key issue in analyzing complex networks. In this paper, a new greedy and hybrid approach based on a community detection algorithm and an MADM technique (TOPSIS) is proposed to cope with the problem, called, ‘Greedy TOPSIS and Community-Based’ (GTaCB) algorithm. The paper concisely introduces community detection and TOPSIS technique, then it presents the pseudo-code of the proposed algorithm. Afterwards, it compares the performance of the solution which is found by GTaCB with some well-known greedy algorithms, based on Degree Centrality, Closeness Centrality, Betweenness Centrality, PageRank as well as TOPSIS, from two aspects: diffusion quality and diffusion speed. In order to evaluate the performance of GTaCB, computational experiments on nine different types of real-world networks are provided. The tests are conducted via one of the renowned epidemic diffusion models, namely, Susceptible-Infected-Recovered (SIR) model. The simulations exhibit that in most of the cases the proposed algorithm significantly outperforms the others, chiefly as number of initial nodes or
probability of infection increases.
\end{abstract}


\section{Introduction}\label{sec:intro}

Social networks permeate our social and economic lives. They play a central role in the transmission of information about job opportunities, and are critical to the trade of many goods and services\cite{Jackson2010}. Such networks also underlie the trade and exchange of goods in non-centralized markets, the provision of mutual insurance in developing countries, research and development, and collusive alliances among corporations, international alliances, and trading agreements; to mention just a few examples\cite{Jackson2003}. Social network analysis focuses on the relationship analysis between social and economic entities with the aim of finding common features between them\cite{AghaMohammadAliKermani2016a}, \cite{Alizadeh2017}. 

One of the most important and attractive research lines in social network analysis, is the analysis of information diffusion in social networks. Diffusion over social networks is a quite common phenomenon, like the spread of rumors, viral marketing of new products, virus propagation and public opinion formation\cite{Alizadeh2014}. Information diffusion is a vast research domain and has expressed research interests of many fields, such as Physics, Biology, etc.\cite{Guille2013},\cite{Kermani2018} \cite{AghaMohammadAliKermani2016}. For example, considering Facebook, where a user Sally updates her status or writes on a friend’s wall about a new show in town that she enjoyed. The information concerning this action is typically passed on to her friends. When some of Sally’s friends make comments on her update, the fact should not be forgotten that the information was flown to the friends of hers and the other users. In this way, the information provided by Sally has the potential to propagate transitively through the network \cite{Chen2013}. 

One of the focal research directions related to information diffusion, is to conduct a study about how to choose individuals (here we call them ‘seeds’) to start the diffusion like that. When the diffusion process terminates, the number of infected individuals in the network can be maximized\cite{Zhao2017}. This problem would be of great importance to many companies as well as individuals that want to do a special promotion of their products, services, and innovative ideas through the powerful word-of-mouth effect (or called viral marketing)\cite{Chen2009}. The above problem, called influence maximization, was first formulated as a discrete optimization problem in\cite{Kempe2003}: A social network is modeled as a graph with nodes representing individuals and edges as connections or relationships between two nodes. Influences are propagated in the network according to an influence model. Given a social network, an influence model and a small number as k, and the influence maximization problem is to find $K$ nodes in the graph such that under the given influence model, the expected number of nodes activated by the $K$ seeds is the largest possible \cite{Chen2010}. 

This problem is generally classified into two separated classes\cite{AghaMohammadAliKermani2017}; competitive class and non-competitive class. In The first class, there are at least two decision makers desiring to spread their information onto a given network and attract as more nodes as possible to their favorite attitudes (such as \cite{AghaMohammadAliKermani2017} and \cite{Carnes2007}). In the non-competitive case, however, only a single decision maker is present (such as \cite{Shang2017}). 
In this work, the focus is on the non-competitive case in which there is a decision maker attempting to spread a piece of information on a social network as much as possible. In this paper, a new algorithm to solve the influence maximization problem is proposed with three main features: (1) The proposed algorithm takes the topological features of the nodes into account. (2) The nodes’ features are taking into consideration simultaneously by making use of one of the multi attribute decision making methods. (3) The fact that the social networks hold some communities is strictly true.

So, the proposed algorithm in the present paper is trying to find the most influential nodes in a given network based on multiple criteria (multiple centrality measures). In the existing works, a number of centrality measures have been proposed to identify the influential nodes \cite{Borgatti2006}. However, all of them focused on only one centrality measure and they have some limitations and disadvantages \cite{Opsahl2010}. Meng et al. showed that the centrality measures have different performance to find the influential nodes\cite{Meng2017}. If only one centrality measure is adopted, then the rankings of identifying the influential nodes may be different by using a different centrality measure\cite{Du2014}. So, to cope with this inefficiency in finding the most influential nodes in social networks, one of the multi attribute decision making techniques (TOPSIS) is utilized in the present work.

On the other hand, since the social networks have a feature that they are community-based \cite{Newman2006}, we come to the conclusion that by exploring the community structures naturally embedded in a social network, efficient algorithms can be developed to address the influence maximization problem \cite{Chen2014}. 
Since there are no algorithms in which a MADM method and community detection algorithm are utilized in the influence maximization simultaneously, the main novelty of the present work is to consider the multiple centrality measures to find the most influential nodes in each community to find the best seed nodes to maximize the influence on the social networks.

The rest of the paper is organized as follow: \Cref{sec:lit} deals with the literature review of influence maximization problem. The third section of the paper defines the considered problem. The suggested algorithm and the used methods are elaborated in the fourth one. The following part expresses the efficiency of the proposed algorithm using some real well-known datasets in the literature. 

\section{Literature Review}
\label{sec:lit}
The influence maximization problem has been proposed and studied by Domingos and Richardson in 2001 \cite{Domingos2001} and was followed by Kempe et al. in 2003 \cite{Kempe2003}. This problem can be mathematically defined on a network $G=(V,E)$, under one of the influence models such as independent cascade or linear threshold. Let $n$ and $u$ be the numbers of elements of $V$ and $E$, respectively. Let $K$ be a positive integer with $n>K$. So, the influence maximization problem is finding a set $A_K^*$ of $K$ nodes to target for initial activation such that $\sigma(A_K^*)\geq \sigma(S)$ for any set $S$ of nodes, i.e.
\begin{equation}\label{eq:1}
    A_K^* = argmax_{A\in \{S \subset V;\vert S\vert =K \}} \sigma(A)
\end{equation}
Where $\vert S\vert $ stands for the number of elements of set $S$ and $\sigma(S)$ stands the number of infected nodes in termination of information diffusion if $S$ is the set of initial nodes for information diffusion process.

In this section, at first, some of the well-known influence models used in previous works are reviewed. The next subsection reviews some of the classic works. The third part of this section is dealing with some of works which investigate the problem from a multi attribute point of view. And finally, the papers in which a community-based approach is adopted to cope the problem are reviewed.

\subsection{Influence models} \label{sec:2_1}
Let the diffusion of information on social networks proceeds along discrete time phases ($t=0,1,2,\dots$). Each node ($v\in V$) can be either active (infected) or inactive (non-infected). The influence models try to describe the status of the nodes based on different initial nodes sets in each step. In this subsection, we review four main and well-known influence models, named, Independent Cascade (IC), Linear Threshold (LT), Susceptible-Infected-Recovered (SIR) and Susceptible-Infected-Susceptible (SIS). It should be noted that IC and LT are known as stochastic diffusion models and SIS and SIR are known as epidemic model in the literature \cite{Chen2013}. 

The IC model considers the network as $G=(V,E)$, the influential probability $p(.)$ on all arcs, and an initial seed set $S_0$ as the model’s inputs. Then it will generate the active set of nodes in each time phase ($S_t$) by the following rule. In each step ($t\geq 1$), the first set $S_t$ to be $S_{t-1}$; next for every inactive node $v\notin S_{t-1}$, for every node $u \notin N^{in} (v)\cap( S_{t-1}\backslash S_{t-2})$, $u$ executes an activation attempt by performing a Bernoulli trial with in a probability of success $p (u,v)$; if successful we add $v$ to $S_t$ and $v$ is activated by $u$ at time $t$. 

The LT model considers the network as $G=(V,E)$, the influence probability $w(.)$ on all arcs, and an initial seed set $S_0$ as the model’s inputs. Then it will generate the active set of nodes in each time step ($S_t$) by the following rule. In $t=0$, each node ($v\in V$) independently selects a number at random ($\theta_v$) from a uniform distribution in the range [0,1]. In each step ($t\geq 1$), first set $S_t$ to be $S_{t-1}$; then for any inactive nodes $v\in V\backslash S_{t-1}$, if the total weight of the arcs is at least $\theta_v$ from its active in-neighbors, then $v$ will be activated and added to $S_t$.

SIR model of infection is one of the most renowned infection models widely used in social network simulation and analysis, it is also mapped onto a bond percolation model \cite{Newman2002}, \cite{Newman2003},in which each individual (or node) transitions between several possible states, which typically include state $S$ (for susceptible), state $I$ (for infected), and state $R$ (for recovered or removed). A node in state $S$ has not the disease but is susceptible to get the disease upon contact with an infected node. A node in state $I$ has the disease and can transmit the disease to susceptible nodes upon contact, with infection rate $\beta$, which is interpreted as the probability of successful transmission of the disease from an infected node to a susceptible node in a time unit. Consider the network of relationship as $G=(V,E)$, where $S_t$ is the set of susceptible nodes, $I_t$ denotes the set of Infected nodes and $R_t$, the set of Recovered ones at time $t$; here if $v\in S_t$ , it becomes infected in $t+1$ if it has at least one neighbor $u\in I_t$, who successes to spread the infection on $v$, with Bernoulli success probability of $p=w_{v,u}×\alpha$ ; in which $w_{v,u}\in [0,1]$ represents the weight of arc connecting $u$ to $v$ where $\alpha\in [0,1]$ is the infectiousness rate of $u$ at time $t$. In such a way:
\begin{equation}
    v\in \begin{cases}
     I_{t^\dagger}, & \text{if}\quad t\leq t^\dagger\leq t+L \\
    R_{t^\dagger}, & \text{if}\quad t^\dagger>t+1+L
    \end{cases}    
\end{equation}
The process ceases whenever the network experience an steady state where all infected nodes are recovered \cite{Hethcote1976}.

\subsection{Classical influence maximization}\label{sec:2_2}
Kempe et al. proved that the influence maximization problem is NP-hard, and proposed a greedy approximate algorithm considering LT, IC and WIC , which guarantees that the influence spread is within (1-1⁄e) of the optimal solution \cite{Kempe2003}. They also showed in experiments that their greedy algorithm significantly outperformed the classic degree and centrality-based heuristics in influence spread. Afterwards, there were many studies which proposed different algorithms to find the best set of initial nodes with influence spread. 

For example, Leskovec et al. \cite{Leskovec2007} proposed an improved greedy algorithm by introducing a “Cost-Efficient Lazy Forward” (CELF) scheme. The CELF algorithm can speed up the greedy algorithm by 700 times. Following the kempe et al’s work\cite{Kempe2003}, Goyal et al.\cite{Goyal2011} optimized CELF by exploiting sub-modularity and experiments and proposed CELFpp algorithm. They showed that CELFpp algorithm is 35\% $\sim$ 55\% faster than CELF. In another research, Chen et al.\cite{Chen2009} developed the New-Greedy and Mixed-Greedy algorithms to improve the greedy algorithm in different ways. Liu et al. proposed a bound linear approach to influence computation and influence maximization\cite{Liu2014}. Kermani et al.\cite{AghaMohammadAliKermani2016} proposed a multi objective mathematical program with linear objectives and constraints in search of the seed nodes in social networks. Since the proposed model was solved by an exact algorithm (CPLEX), they claimed that their model finds the optimal solution for influence maximization problem. 

Additionally, some heuristic algorithms have been proposed to cope with the influence maximization problem. These approaches are trying to find the top-k nodes in social networks based on degrees or other centrality measures. Recently, some studies have been made in which some meta-heuristic based algorithms were proposed to deal with the influence maximization problem. For example, Yang and Weng \cite{Yang2012} proposed the swarm intelligence-based algorithm (ant-colony optimization algorithm) to consider the influence maximization problem. The proposed algorithm was evaluated using a co-authorship data set and the obtained experimental results showed that the proposed algorithm outperforms two well-known benchmark heuristics. Other metaheuristic algorithms such as genetic algorithm \cite{Bucur2016}, simulated annealing algorithm \cite{Jiang2011}, particle swarm optimization algorithm \cite{Gong2016} and cuckoo search algorithm \cite{Gandomi2013} have been utilized to deal with the influence maximization problem, too. 

\subsection{Multi-attribute-based approach for influence maximization}
Another wave in influence maximization research line is to find the influential nodes (key nodes) in consideration of more than one criterion. For the first time, Mesgari et al. in 2013, presented a novel approach in a conference in Bielefeld,  in which they utilized TOPSIS method to find the key nodes in social networks \cite{Mesgari2015}. They examined their approach using three datasets in various sizes and sub-structures. In another study, Fox and Everton applied a hybrid method based on AHP and TOPSIS to find the influential nodes in Noordin Dark Network. Additionally, they discussed a bit about the sensitivity of the nodes’ rank based on changes in weights of the criteria \cite{Fox2014}. Zhang et al. studied the problem of nodes’ importance in the research and the developmental team \cite{Zhang2013}. They considered the eight criteria to identify the importance of the nodes; there were four criteria from centrality measures (degree centrality, betweenness centrality, closeness centrality, and eigenvector centrality) and the others from structural holes of complex networks (effective size, efficiency, constraint, and hierarchy). They used a Fuzzy AHP method to identify the weights of the mentioned criteria and then a TOPSIS method to rank the nodes based on their importance. Du et al. proposed a TOPSIS method to identify the influential nodes in social networks \cite{Du2014}. They applied different types of centrality measures as TOPSIS’s attributes in different networks. The effectiveness of the proposed method considering SIS model as the influence model, is examined by comparing the results with some of the benchmark methods. The main weakness of Du et al.’s work was the consideration of the same weights for TOPSIS’s attributes in nodes rank. The proposed method has improved in \cite{Du2014}. The authors proposed a new algorithm to calculate the weight of each attribute. In order to evaluate the performance of the method, they used the SIS model as the influence model to simulate the diffusion process in four real networks.

\subsection{Community-based approaches for influence maximization}
Another wave in influence maximization research line is developing community-based approaches to solve the problem. Community structure is defined as the division of network nodes into groups, within which nodes are densely connected while between which they are sparsely connected \cite{Jaquet2009}. The main reason of utilizing community-based approach to this problem is the reduction of the computational time and an increase in the performance.

Cao et al. proposed the first community-based influence maximization algorithm OASNET (Optimal Allocation in a Social NETwork). They transformed the influence maximization problem into an optimal resource allocation problem. Also, they assumed that different communities are independent of each other and influence cannot spread across different communities \cite{Cao2010} .Then, they proposed a recursive relation to find the influential nodes in social networks. Zhang et al. studied the problem of influence maximization on networks with community structure. The authors constructed an information transfer probability matrix from the weighted network \cite{Zhang2013a}. Then they applied the k-medoid clustering algorithm to identify the Top-K (influential) nodes. The performance of the proposed method has been investigated using LFR synthetic networks \cite{Lancichinetti2008} and several real-world network. Wang et al. proposed an algorithm called Community-based Greedy Algorithm for mining top-K influential nodes in social networks \cite{Wang2010}. Their empirical studies show that their method is faster than the state-of-the-art Greedy algorithm to find the influential nodes. Chen et al.\cite{Chen2014} developed a new framework to tackle the influence maximization problem with an emphasis on time efficiency. The proposed framework consists of three phases; community detection, candidate generation and seed selection. They tested the proposed framework’s efficiency and scalability on both synthetic and real datasets and showed the developed framework outperforms the state-of-the-art algorithms. One of the most recent studies in this field is Shang et al.’s research \cite{Shang2017}. To solve the influence maximization problem, they proposed a new algorithm named as CoFIM. The developed algorithm contains two phases; seeds expansion and intra-community propagation. The first phase is the expansion of seed nodes among different communities at the beginning of the diffusion. The second phase is the influence propagation within communities which are independent of each other. To evaluate the performance of the proposed algorithm with state-of-the-art algorithms, they used some synthetic and real-world large datasets.

\section{Problem Definition}
We consider a directed social network $G= (V,E)$, with $\vert V\vert =n$ and $\vert E\vert =u$ with the edge weights $w_{{ij}}$ between nodes $i$ and $j$. In this paper, we assume that the influence model is SIR and the time proceeds in discrete periods. Furthermore, based on the mainstream of the influence maximization literature, it is assumed that the model is progressive. We assume that at time $t = 0$, the seed nodes are active. An active node stays infectious for $L$ periods when she can infect its immediate neighbors with given infectiousness rates, and be \emph{‘Recovered’} after that, so that she does not have a chance to infect others any more. Hence, the process lasts, when there is at least one active node, and terminates immediately when the last active nodes are Recovered. The focus of this paper is to find a set of seeds maximizing the number of infected nodes at the cessation of the infection process.

\section{The Proposed Algorithm}\label{sec:4}
In this paper, a new algorithm is proposed to cope with the influence maximization problem given the SIR influence model. The proposed algorithm is called \emph{“Greedy TOPSIS and Community Based”} algorithm which is abbreviated to GTaCB. The algorithm employs two well-known methods in community detection literature and multi-attribute decision making. The procedure utilized the methods and the result of the implementation of the proposed algorithm on a small dataset being introduced in \cref{sec:4_1}.

\subsection{Procedure}\label{sec:4_1}
\begin{figure}
    \centering
    \includegraphics[width=\textwidth]{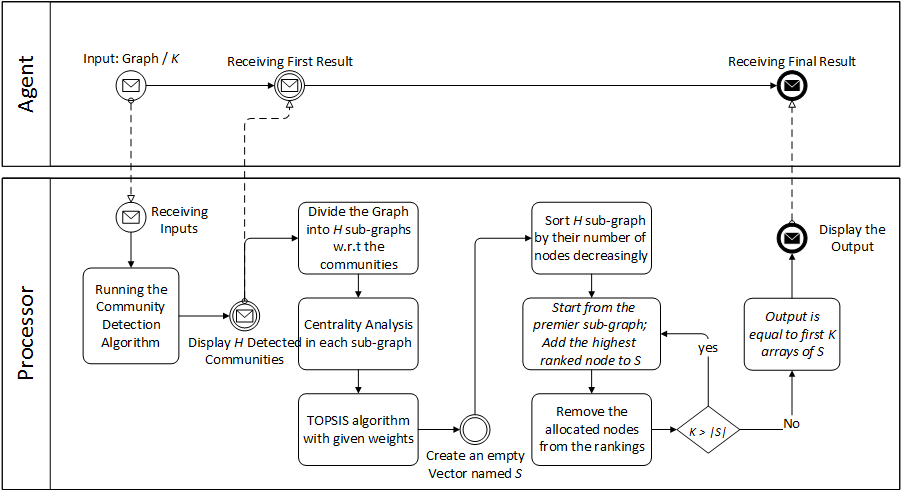}
    \caption{The schematic process of GTaCB algorithm}
    \label{fig:1}
\end{figure}

The schematic procedure of the proposed algorithm is illustrated in \cref{fig:1}. First, GTaCB runs a community detection algorithm to find $K$ partitions within the graph, which is explained in \ref{sec:4_1_1}. In some cases, however, it is possible that the graph doesn’t contain $K$ distinct communities with respect to the community detection algorithm used. Therefore, given it detects H communities, GTaCB divides the whole graph, into H sub-graphs with respect to the communities. Afterwards, in each sub-graph the Centrality Analysis and then TOPSIS technique has to be conducted, which is explained in \ref{sec:4_1_2}. Finally, to select the set of seed nodes as the output of the algorithm, it sorts the sub-graphs by their number of nodes, decreasingly. Then, as a loop, it starts from the premier sub-graphs and allocates the highest ranked node of each one to the set of seeds, $S$, and removes the allocated nodes from the rankings. In the case that $K>H$, for the remaining desired seeds, the algorithm uses the next ranked nodes in each iteration until it is satisfied.

\subsubsection{Community Detection}\label{sec:4_1_1}
Detecting the communities in networks is a big challenge for which many methods and algorithms have been proposed in the last decades, within different scientific disciplines such as Physics, Biology, Computer Science and Social Sciences\cite{Lancichinetti2008}. There are different algorithms identifying the communities each of which are used for different types of networks, depending on the network features and the characteristics of the community detection algorithm\cite{Orman2011}, \cite{Karimi-Majd2015}.

Since the proposed algorithm in this paper is trying to find $K$ best influential nodes in $K$ communities of social networks, it is preferred to use a community detection algorithm in which the number of communities can be tuned before running the algorithm. Graph Community detection by Spectral Clustering (GCSC) algorithm proposed by Hespanha\cite{Hespanha2004} is a community detection in which the number of desired communities is one of its inputs. So, we implement this algorithm to divide the graph into the partitions in a way that minimizes the edge-costs of each partition.

Let $G=(V,E)$ represent a directed graph, where $V$ is the set of vertices with edge set of $E$, and there are $k$-partitions (subsets) of $V$ denoted as $P=\{V_1,V_2,\dots,V_k\}$ where $V_i \cap V_j=\emptyset$ $;i\ne j\in{1,\dots,k}$ and $V_1\cup V_2\cup,\dots,\cup V_k=V$
An edge-costs function of graph partitioning $P$ can be defined as

\begin{equation}
    C(P)=\sum_{i\ne j} \sum_{{(v,\Bar{v})\in E; v\in V_i , \Bar{v}\in V_j}} c(v,\Bar{v}); \quad c:E \rightarrow [0,\inf)
\end{equation}

\subsubsection{Multi Attribute Decision Making}\label{sec:4_1_2}
As it is illustrated at \cref{fig:1}, the process of identifying the influential nodes is initiated by the community detection algorithm and then if the desired number of partitions is found, the processor inserts the detected partitions into the algorithm calculating centrality measures. Then TOPSIS (or any other MADM tool) helps us select the top scored node of each partition.

Hwang and Yoon \cite{Hwang1981} introduced TOPSIS (Technique for Order Preference by Similarity to an Ideal Solution) which has become one of the most prevalent MADM (Multi-Attribute Decision Making) methods in the literature. In such a method, there is a finite set of alternatives about to be evaluated and ranked by the criteria or attributes which are individually weighted before. TOPSIS has been utilized in many research areas such as Supply Chain Management, Facility Location, HSE and Project Management. In our algorithm, we defined the criteria as network centrality measures. TOPSIS suggests herewith the alternative which is the closest to the ideal solution and the farthest from negative ideal solution as the most influential node in each community \cite{Wei2010}. TOPSIS has following steps \cite{Mesgari2015}:
\begin{itemize}
    \item Step1: Creating a decision matrix with $m$ rows as alternatives (nodes) and $n$ columns for criteria: $X= \begin{bmatrix}
x_{11} & \cdots & x_{1p} \\
\vdots & \ddots & \vdots \\
x_{m1} & \cdots & x_{mp}
\end{bmatrix}$.
	\item Step2: The normalization step in which X is converted to R by $r_{ij}=x_{ij} (\sum_{i=1}^m x_{ij}^2 )^{-1/2}  \forall j$
	\item Step3: Determining the weight normalized matrix $[t_{ij} ]_{m×p}=[w_j\quad r_{ij} ]_{m×p}$ showing the relative importance of each criterion.
	\item Step4: Let $J_+$ represent the set of benefit criteria and $J_-$ as the set of cost criteria, determine the positive ideal solution ($t_j^+$) and the negative ideal solution ($t_j^-$) as:
    \begin{itemize}
        \item $t_j^+=\{(\max t_{ij} \vert i=1,\dots,m ; \forall j\in J_+ ),(\min t_{ij}\vert i=1,\dots,m ; \forall j\in J_- )\}$
        \item $t_j^-=\{(\max t_{ij}\vert i=1,\dots,m ; \forall j\in J_- ),(\min t_{ij}\vert i=1,\dots,m ; \forall j\in J_+ )\}$
    \end{itemize}
	\item Step5: Calculating the distance of each alternative from its positive and negative ideal solution by: $S_i^+=\sqrt{(\sum_{j=1}^p (t_{ij}-t_j^+ )^2}; \quad S_i^-=\sqrt{(\sum_{j=1}^p t_{ij}-t_j^- )^2};\quad i=1,\dots,m$
	\item Step6: Determining the relative closeness to the ideal solution for all nodes, as: $C_i^*=\frac{S_i^-}{S_i^-+S_i^+}$
	\item Step7: At the final step, the nodes are ranked based on $C_i^*$ values, in descending order.

\end{itemize}

Based on the previous works using TOPSIS methods to identify the most influential nodes in social networks, we considered Degree Centrality (DC), Closeness Centrality (CC), Betweenness Centrality (BC) and PageRank (PR) as the attributes in this decision-making process\cite{Du2014}, \cite{Mesgari2015}, \cite{Hu2016}. 

\subsubsection{The GTaCB algorithm}\label{sec:4_1_3}
So, the identified influential set of seeds consist of all highest ranked nodes of $k$-partitions ($P$). in the remainder of the section, firstly, we define the main notations used in the paper in Algorithm~\ref{algorithm:1}, secondly, the pseudo-code of the proposed algorithm is exhibited in \cref{fig:7} and then the paper gives an example on a small network and compares the proposed algorithm results with those of TOPSIS.

\begin{table}[]
\label{tab:1}
\caption{Main notations}
\begin{tabular}{ll}
\toprule
Parameter & Definition                                                            \\
\midrule
$n$         & Number of nodes in the network                                        \\
$u$         & Number of edges between nodes in the network                          \\
$V$         & Set of $n$ nodes                                                        \\
$E$         & Set of $u$ edges                                                        \\
$w_{ij}$         & Weight of the edge connecting $i$ to $j$                                  \\
$K$         & Number of initial seeds                                               \\
$S^t$         & Set of susceptible nodes at time $t$                                    \\
$I^t$         & Set of infected nodes at time $t$                                       \\
$R^t$         & Set of recovered nodes at time $t$                                      \\
$p^t_{ij}$         & Probability of transmission the infection/influence from $i$ to $j$, if $i\in I^t$ and $j\in S^t$  \\
$L$         & Number of periods a node stays infected                               \\
$\alpha_r$     & Infectiousness rate in rth period of infection; $r = 1,\dots,L$            \\
$k$         & Relative infectiousness                                               \\
$\Gamma$     & Number of infected nodes at the end of the infection process          \\
$\tau$       & Length of the infection process (Number of periods the process lasts) \\
$\mu$        & Diffusion speed\\
\bottomrule
\end{tabular}
\end{table}

With using the notations described in \cref{tab:1}, the paper proposes GTaCB as follow:

\begin{algorithm}[H]
\label{algorithm:1}
\caption{GTaCB pseudo-code}
  \DontPrintSemicolon
  \KwInput{$V, E, G, K, W$}
    \textbf{Initialization}: $S=zeros(1,k)$; \quad \textcolor{teal}{\# an empty vector of seeds}\;
    \textbf{Community Detection}: $P = GCSC(G,K)$; \quad \textcolor{teal}{\# returns $P$ as the set of $K$ Partitions}\;
    \textbf{Sorting the Communities}: sort vector P w.r.t the number of nodes, decreasingly.\;
    \textbf{Seed Detection}: \;
    \For{$p = 1:length(P)$}{
        $P_p=$the subgraph of $G$, containing vertices in $p^{th}$ subset of $P$;\;
        \textbf{Centrality Measures Calculation}: $C = Centralities(P_p)$;\;
        \textcolor{teal}{\# returns the matrix of $C$ containing each node’s centrality values in subgraph $P_p$}\;
        \textbf{TOPSIS Calculation}: $T = TOPSIS(C,W)$;\;
        \textcolor{teal}{\# returns nodes of $P_p$  as the vector $T$ in descending order by their TOPSIS ranks}\;
        \eIf{$p\leq K-\lfloor K/length(P)\rfloor \times length(P)$}{
            $S=[S,T(1, 1 : \lceil K/length(P) \rceil$];\quad \textcolor{teal}{\# allocates the best ranked nodes to the Seeds}\;
        }{
            $S=[S,T(1, 1 : \lfloor K/length(P)\rfloor$];\quad \textcolor{teal}{\# allocates the best ranked nodes to the Seeds}\;
        }
    }
\end{algorithm}

\begin{figure}
    \centering
    \includegraphics[width=.9\textwidth]{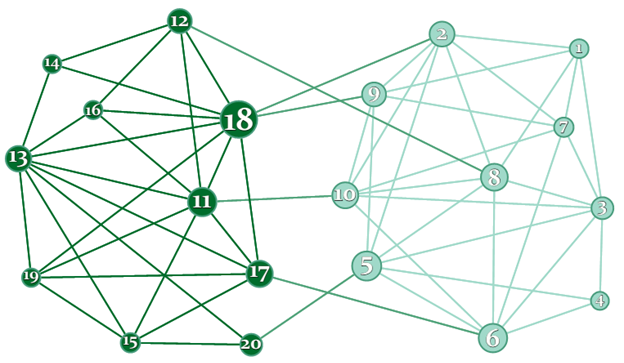}
    \caption{Ex1 with two communities}
    \label{fig:2}
\end{figure}

Let us consider a small synthetic undirected network named ‘Ex1’ having 20 nodes, with equal weight values on its edges. The network has been generated by ‘Random Modular Graph’\footnote{\url{http://strategic.mit.edu/downloads.php?page=matlab\_networks}} algorithm programmed by MIT Strategic Engineering Research Group, the inputs of which are $n=20$, $c=2$, $p=0.3$ and $r=0.9$. The graph depicts in \cref{fig:2} with distinct colors for its communities.

Consider if there are only two seeds to start the infection with. According to their centrality measures (DC, CC, BC and PR), \cref{tab:2} contains the values each node has in the network. Hence, the top two nodes of each centrality’s ranking are selected and highlighted in the table.

\begin{table}[]
\label{tab:2}
\caption{The ranks of the nodes by each centrality (highlighted rows are selected)}
\begin{tabular}{lllllllll}
\toprule
\multirow{2}{*}{Rank} & \multicolumn{2}{l}{BC} & \multicolumn{2}{l}{CC} & \multicolumn{2}{l}{DC} & \multicolumn{2}{l}{PR} \\
                      & NODE     & SCORE       & NODE     & SCORE       & NODE      & SCORE      & NODE     & SCORE       \\
\midrule
\textbf{1}  & \textbf{18} & \textbf{57.55433} & \textbf{18} & \textbf{0.032258} & \textbf{18} & \textbf{9} & \textbf{18} & \textbf{0.07376}  \\
\textbf{2}  & \textbf{5}  & \textbf{34.1868}  & \textbf{2}  & \textbf{0.03125}  & \textbf{11} & \textbf{8} & \textbf{13} & \textbf{0.068025} \\
3                     & 11       & 32.39545    & 6        & 0.030303    & 13        & 8          & 11       & 0.065967    \\
4                     & 6        & 32.22749    & 9        & 0.030303    & 2         & 7          & 5        & 0.058792    \\
5                     & 8        & 27.24242    & 10       & 0.030303    & 3         & 7          & 6        & 0.057705    \\
6                     & 17       & 25.7868     & 11       & 0.030303    & 5         & 7          & 3        & 0.057703    \\
7                     & 10       & 24.51558    & 17       & 0.030303    & 6         & 7          & 8        & 0.057076    \\
8                     & 13       & 23.48831    & 8        & 0.029412    & 8         & 7          & 2        & 0.05656     \\
9                     & 2        & 21.3026     & 5        & 0.028571    & 10        & 7          & 10       & 0.056332    \\
10                    & 12       & 18.68918    & 13       & 0.028571    & 7         & 6          & 17       & 0.050703    \\
11                    & 9        & 18.33593    & 12       & 0.027778    & 9         & 6          & 9        & 0.049483    \\
12                    & 3        & 12.13355    & 3        & 0.027027    & 17        & 6          & 7        & 0.04941     \\
13                    & 20       & 11.72641    & 7        & 0.025641    & 1         & 5          & 15       & 0.044682    \\
14                    & 7        & 4.307143    & 19       & 0.025641    & 12        & 5          & 12       & 0.04442     \\
15                    & 15       & 4.200866    & 20       & 0.025641    & 15        & 5          & 19       & 0.043488    \\
16                    & 1        & 2.383333    & 15       & 0.02439     & 19        & 5          & 1        & 0.042314    \\
17                    & 19       & 1.357143    & 16       & 0.02439     & 16        & 4          & 16       & 0.036251    \\
18                    & 14       & 1.083333    & 1        & 0.02381     & 4         & 3          & 20       & 0.029452    \\
19                    & 16       & 1.083333    & 14       & 0.023256    & 14        & 3          & 14       & 0.029237    \\
20                    & 4        & 0           & 4        & 0.022222    & 20        & 3          & 4        & 0.028641  \\
\bottomrule
\end{tabular}
\end{table}

According to the steps of the proposed algorithm, the communities are detected and shown in \cref{tab:3}, so the graph is split into two sub-graphs, in each of which the centralities and subsequently the TOPSIS scores are calculated independently. So, in community \#1, node 3 is chosen and in community \#2, node 13. However, nodes 18 and 11 would be selected as the first and the second ones ranked as if the graph were not separated.

\begin{table}[]
\label{tab:3}
\caption{The ranks of the nodes via TOPSIS and GTaCB (highlighted rows are selected)}
\begin{tabular}{llllllll}
\toprule
\multirow{2}{*}{Rank} & \multicolumn{2}{l}{TOPSIS}      & \multirow{2}{*}{Rank} & \multicolumn{3}{l}{GTaCB}                &               \\
                      & NODE        & SCORE             &                       & NODE        & Scores     & \multicolumn{2}{l}{Community} \\
\midrule
1                     & \textbf{18} & \textbf{1}        & 1                     & \textbf{3}  & \textbf{1} & \multicolumn{2}{l}{1}         \\
2                     & \textbf{11} & \textbf{0.617439} & 1                     & \textbf{13} & \textbf{1} & \multicolumn{2}{l}{2}         \\
3                     & 5           & 0.609788          & 3                     & 5           & 0.880983   & \multicolumn{2}{l}{1}         \\
4                     & 6           & 0.583636          & 4                     & 6           & 0.571812   & \multicolumn{2}{l}{1}         \\
5                     & 8           & 0.51454           & 5                     & 11          & 0.530511   & \multicolumn{2}{l}{2}         \\
6                     & 13          & 0.505129          & 6                     & 18          & 0.511357   & \multicolumn{2}{l}{2}         \\
7                     & 10          & 0.479036          & 7                     & 7           & 0.486803   & \multicolumn{2}{l}{1}         \\
8                     & 17          & 0.464143          & 8                     & 10          & 0.470492   & \multicolumn{2}{l}{1}         \\
9                     & 2           & 0.441045          & 9                     & 8           & 0.461369   & \multicolumn{2}{l}{1}         \\
10                    & 9           & 0.365922          & 10                    & 2           & 0.446429   & \multicolumn{2}{l}{1}         \\
11                    & 3           & 0.335177          & 11                    & 9           & 0.30307    & \multicolumn{2}{l}{1}         \\
12                    & 12          & 0.332455          & 12                    & 1           & 0.302609   & \multicolumn{2}{l}{1}         \\
13                    & 7           & 0.217104          & 13                    & 15          & 0.284238   & \multicolumn{2}{l}{2}         \\
14                    & 20          & 0.179894          & 14                    & 17          & 0.220245   & \multicolumn{2}{l}{2}         \\
15                    & 15          & 0.16514           & 14                    & 19          & 0.220245   & \multicolumn{2}{l}{2}         \\
16                    & 19          & 0.148921          & 16                    & 12          & 0.167151   & \multicolumn{2}{l}{2}         \\
17                    & 1           & 0.144904          & 17                    & 16          & 0.161097   & \multicolumn{2}{l}{2}         \\
18                    & 16          & 0.081011          & 18                    & 14          & 0.093239   & \multicolumn{2}{l}{2}         \\
19                    & 14          & 0.021391          & 19                    & 4           & 0          & \multicolumn{2}{l}{1}         \\
20                    & 4           & 0                 & 19                    & 20          & 0          & \multicolumn{2}{l}{2}        \\
\bottomrule
\end{tabular}
\end{table}

\begin{figure}
    \centering
    \includegraphics[width=.5\textwidth]{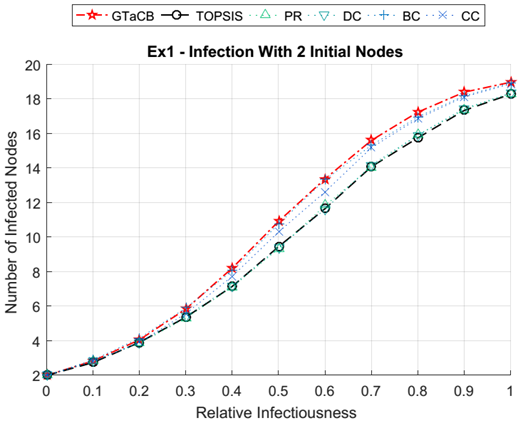}
    \caption{Ex1 SIR infection results with 2 seeds}
    \label{fig:3}
\end{figure}

\cref{fig:3} exhibits the results of an iterative simulation on Ex1 for all selected seeds which are inputted into an SIR model, where maximum iteration is 2000, $L=2$ and $\alpha=(0.3,0.15)$ and the probability of infection is $p_{ij}=k\alpha w_{ij}$.

\begin{figure}
    \centering
    \includegraphics[width=.5\textwidth]{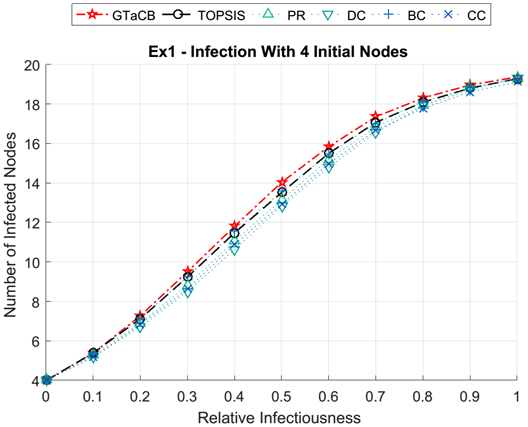}
    \caption{Ex1 SIR infection results with 4 seeds}
    \label{fig:4}
\end{figure}

The results shown in \cref{fig:3} and \cref{fig:4} show that the seeds that GTaCB suggests, infect more nodes through the iterations, particularly than TOPSIS.

\section{Experimental Results}\label{sec:5}

In order to evaluate the performance of GTaCB algorithm in comparison with some other famous approaches, we have simulated the spread of influence of chosen initial nodes detected by each measure, using an SIR model \cite{Jaquet2009}, an epidemic model of infection. All the codes have been written in MATLAB R2016a, and the results are computed at a Windows 8.1 OS with Core i7 Intel CPU of 3.1GHz and 8GB memory.

\subsection{Real-World Datasets}\label{sec:5_1}
To achieve a kind of comprehensive comparable result, it is worth employing an adequate variety of datasets exhibiting satisfactory structural features of real networks \cite{Kempe2003}. Therefore, in this experiment, there are nine different real-world networks on which we have tested the spread of selected seeds’ influence. These datasets are introduced as follows:

\begin{itemize}
    \item Abrar: A network of SMS connections between university students in industrial engineering and computer engineering at a higher education institute in Tehran named ‘Abrar’, between the years 2010 and 2011 \cite{AghaMohammadAliKermani2014}.
\item USAir: The network of North American Transportation Atlas Data (NORTAD) contains geographic data sets for transportation facilities in Canada, Mexico, and the United States. \url{http://vlado.fmf.uni-lj.si/pub/networks/data/default} 
\item EuroSiS: The network based on collaborations between “European Science in Society” and agents, realized in a WebAtlas study among 12 Countries in Europe, which is accessible at: \url{https://github.com/gephi/gephi/wiki/Datasets} 
\item OCLinks: An online weighted social network created based on students’ online message interactions through an online community at the University of California, Irvine. The weight of an edge is defined as the number of messages sent over a period from April to October 2004. \url{https://github.com/gephi/gephi/wiki/Datasets}
\item Yeast: A dataset of unweighted networks representing protein-protein interactions in budding yeast based on an innovative interactive detective study done by Bu et al. \cite{Bu2003}.
\item Geom: A weighted graph obtained from computational geometry collaborations among authors who had any jointly publishing work. The weights representing number of works each pair of nodes co-authorized published. \url{http://vlado.fmf.uni-lj.si/pub/networks/data/default}
\item HEP-th: Also, is known as “High Energy Collaboration”, is a weighted indirect graph illustrating the posting preprints between physicists in the field of “High Energy Physics” theory E-Print Archive from the beginning of January, 1995 until the last day of 20th century. The graph datasets are accessible at \url{http://www-personal.umich.edu/~mejn/netdata/}
\item PGP-Giant: The network of users who shared confidential information via an encryption algorithm called Pretty-Good-Privacy. These interactions made an edg list of the giant component of this graph in 2004 by Boguñá et al.\cite{Boguna2004} and the dataset is available at: \url{http://deim.urv.cat/~alexandre.arenas/data}
\item Slashdot: A news website which features user-submitted and editor-evaluated currents of primarily technology-oriented news. After 2002, it allows users to tag each other as friends or foes. The network contains friend/foe links between the users of Slashdot in Feb. 2009. The dataset is available at: \url{http://snap.stanford.edu/data/soc-sign-Slashdot090221}

\end{itemize}

\begin{figure}
    \centering
    \includegraphics[width=.95\textwidth]{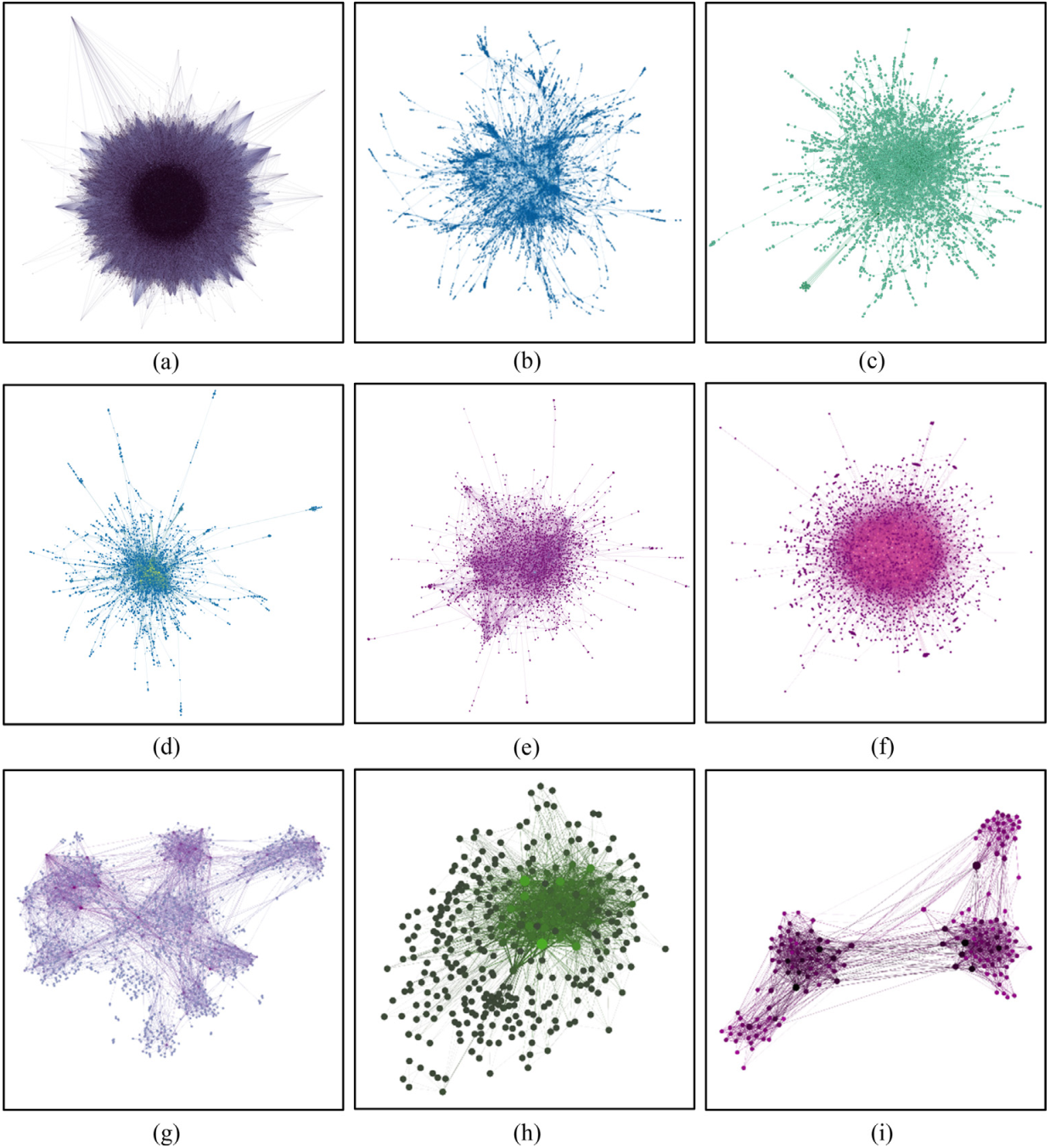}
    \caption{The visualizations of Slashdot(a), PGP (b), HEP-th(c), Geom(d), Yeast(e), OCLinks(f), EuroSiS(g), USAir(h) and Abrar(i) graphs – The layouts are based on Force-Atlas algorithm}
    \label{fig:5}
\end{figure}

The properties of these real-world datasets are introduced in \cref{tab:4}, while \cref{fig:5} illustrates their topography:

\begin{table*}[]
\label{tab:4}
\caption{Properties of real-world datasets}

\resizebox{\textwidth}{!}{\begin{tabular}{llllllllll}
\toprule
                                & Abrar Ins. & US Airlines    & EuroSiS       & OCLinks               & Yeast    & Geom          & HEP-th        & PGP-Giant           & Slashdot   \\
\midrule
Network Type                    & Friend\footnote{Friendship} & Tran.\footnote{Transportation} & Collab.\footnote{Collaboration} & Online\footnote{Online Social Network} & Biology  & Collab. & Coauth.\footnote{Co-authorship} & Inf.\footnote{Information Sharing} & Friend \\
\#Nodes                         & 163        & 332            & 1285          & 1899                  & 2361     & 7343          & 8361          & 10680               & 82140      \\
\#Edges (Directed)              & 3113       & 2126           & 7524          & 20296                 & 14364    & 23796         & 31502         & 48632               & 549202     \\
\#Strongly Connected Components & 1          & 332            & 511           & 601                   & 2361     & 7343          & 8361          & 1                   & 1          \\
Network Diameter                & 5          & 6              & 14            & 8                     & 16       & 12            & 17            & 24                  & 12         \\
Average Path Length             & 2.466      & 2.563          & 4.943         & 3.197                 & 4.647    & 4.006         & 5.16          & 7.485               &            \\
Average Degree                  & 38.196     & 12.807         & 11.711        & 21.375                & 6.084    & 3.241         & 3.768         & 9.107               & 13.372     \\
Maximum Out-Degree              & 51         & 99             & 98            & 237                   & 60       & 101           & 23            & 205                 & 2548       \\
Maximum Betweenness             & 2599.2    & 5286.2        & 162757.1     & 148225.3             & 36248.0 & 46776.0      & 25686.0      & 14959584.7         &   550761.3      \\ \bottomrule
\end{tabular}}
\end{table*}

\begin{algorithm}[H]
\label{algorithm:2}
\caption{SIR Model Simulator}
  \DontPrintSemicolon
  \KwInput{$V, E, G, k, \alpha, L, itermax, K$}
    \textbf{Main Steps}:\\
    infect N(Seeds) as initial seed\\
    $\Psi=zeros(n,itermax)$\\
    \For{$iter = 1:itermax$}{
        Calculate the number of active infected nodes and put it into $C_{inf}$;\\ $t=1$\\
        \While{$C_{inf}>0$}{
            \For{$i=1:K$}{
                $J(i)$=set of infected nodes which have a directed edge to node $i$\\
                \If{$i$ is susceptible and $J(i)\ne \emptyset$}{
                    infect $i$ with probability of $p(i)= k\sum_j w_{ij}\alpha_s, j\in J(i)$
                }
            }
            update the state of all nodes;\\
            update $C_{inf}$;\\
            $t=t+1$
        }
        $\gamma$ = vertical vector of nodes at which infected nodes take 1, and others 0;\\
        $\Psi(:,iter)=\gamma$
    }
    Let $\psi(Seeds)$ be a vector with average of $\Psi$ matrix on iterations
\end{algorithm}

Firstly, we divided these graphs into $K$ clusters via GCSC algorithm as it is explained in \ref{sec:4_1_1}, for instance, \cref{fig:6} depicts the different identified communities in PGP graph:

\begin{figure}
    \centering
    \includegraphics[width=.95\textwidth]{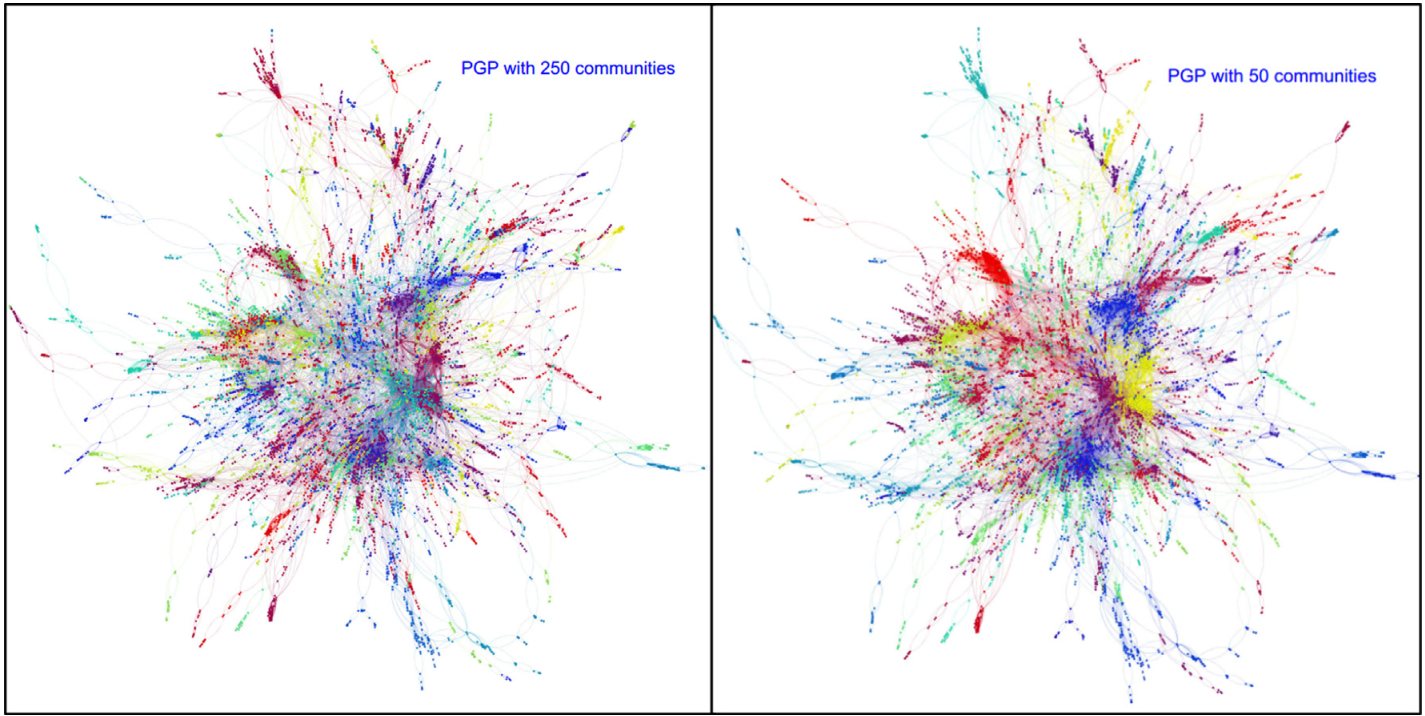}
    \caption{PGP graph with 50 (right) and 250 (left) detected communities. The colors represent different identified communities}
    \label{fig:6}
\end{figure}

If we denote the set of top $K$ nodes which are ranked by $i^{th}$ algorithm as $S_i$, Jaccard Coefficient (JC) between $i^{th}$ and $j^{th}$ algorithms’ detected seeds can be defined as follows,\cite{Jaccard1912}:
\begin{equation}
    JC_{ij}=\frac{\| S_i \cap S_j \|}{\| S_i \cup S_j \|}
\end{equation}

\begin{figure}
    \centering
    \includegraphics[width=.95\textwidth]{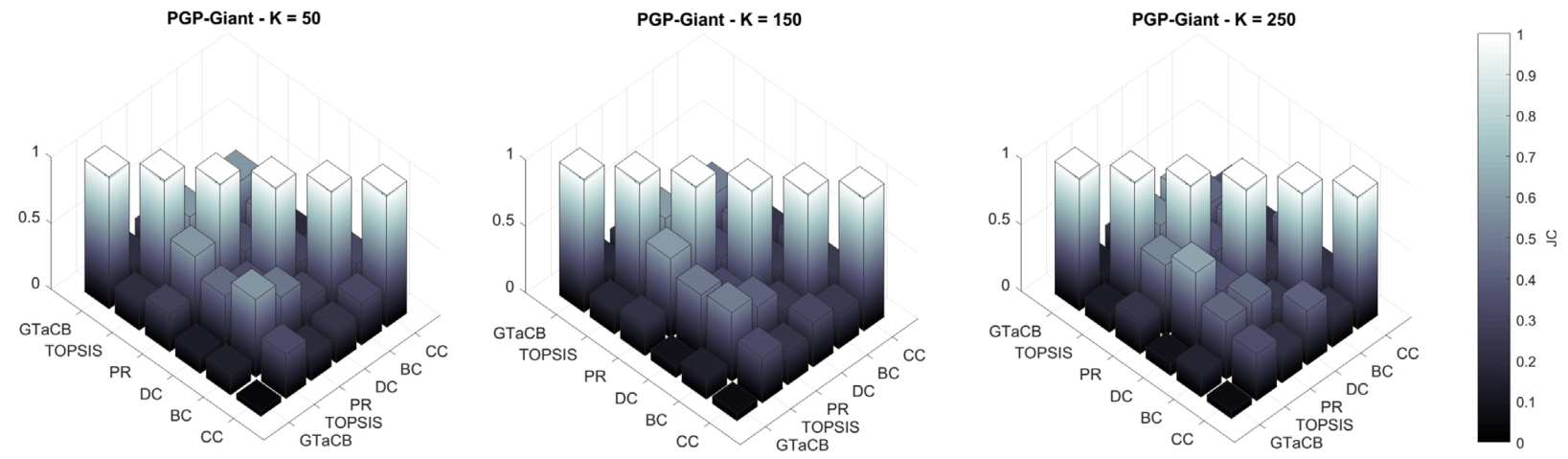}
    \caption{The relationship between each pair of utilized algorithms’ detected seeds in PGP network}
    \label{fig:7}
\end{figure}

\begin{figure}
    \centering
    \includegraphics[width=.95\textwidth]{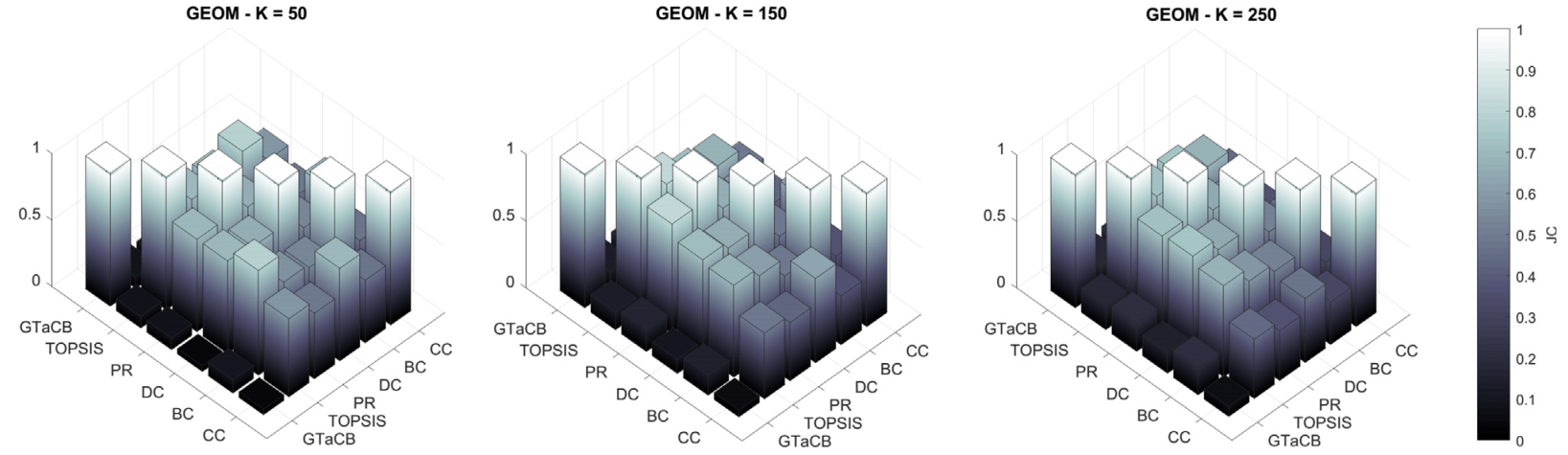}
    \caption{The relationship between each pair of utilized algorithms’ detected seeds in GEOM network}
    \label{fig:8}
\end{figure}

\begin{figure}
    \centering
    \includegraphics[width=.95\textwidth]{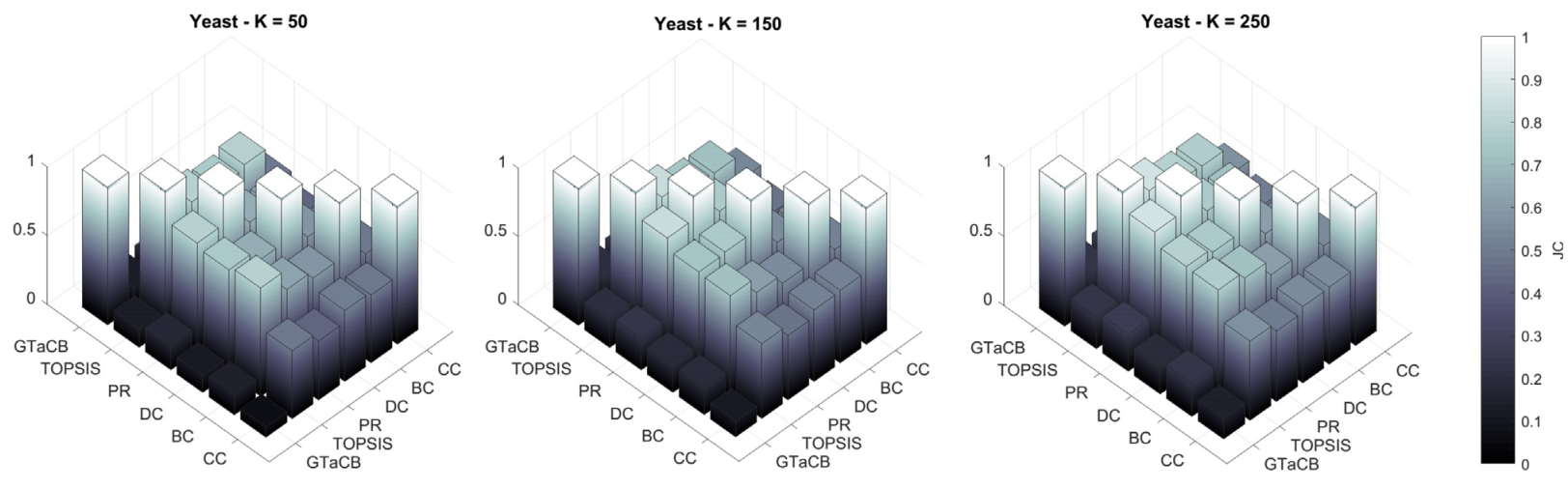}
    \caption{The relationship between each pair of utilized algorithms’ detected seeds in Yeast network}
    \label{fig:9}
\end{figure}

Accordingly, \cref{fig:7}, \cref{fig:8} and \cref{fig:9} show JC values between all pairs of detected seeds by every algorithm in the 3 networks respectively: PGP, GEOM, and Yeast. The figures demonstrate that GTaCB has the least JC in comparison with the others unlike those of TOPSIS’s, whose JC is the highest in terms of centrality. The figures show that the output of TOPSIS has the commonest nodes with the utilized algorithms, however it would not be unanticipated due to the inherency of an MADM technique. It can be clearly seen in our tests that GTaCB’s relationships with PR and BC were higher than its relationship with CC or the rests, especially as $K$ increases. It also demonstrates that the majority of individual nodes, selected by the proposed algorithm, are not considered as “influential” by well-known centralities.

\subsection{SIR model results}\label{sec:5_2}
In order to achieve a quantitative analysis of the proposed algorithm, a modified SIR model of infection is utilized, to which Jaquet and Pechal \cite{Jaquet2009} have imparted a new parameter named ‘relative infectiousness’ that multiplies in all values of infectiousness rate vector ($\alpha_i$), it helps us extract a large spectrum of infection conditions by increasing this parameter from 0 to 1. The Pseudo-Code of this simulation is written in the Algorithm~\ref{algorithm:2}.

The top $K$ ranked nodes of each algorithm ($Seeds_{i.K}$) is inputted to Algorithm~\ref{algorithm:1} as the initial nodes. Parameter settings were the same for all of our employed networks, where $L=2$ and $\alpha=(0.30  0.15)$, and the number of iterations was varied between 100 and 500 in each network (since the process was too long for larger networks), it guarantees an evenhanded conclusion. All the results were stored and for this paper 1,445,000 times Algorithm~\ref{algorithm:1} has been called to simulate the result on the whole. For each case, we have collected two simulated outcomes:
\begin{itemize}
    \item The average number of infected nodes through the iterations($\Gamma$).
    \item The average number of periods it lasted to get a steady step in the iterations ($\tau$).
\end{itemize}

\subsection{Diffusion Quality of GTaCB}
\begin{figure}
    \centering
    \includegraphics[width=.95\textwidth]{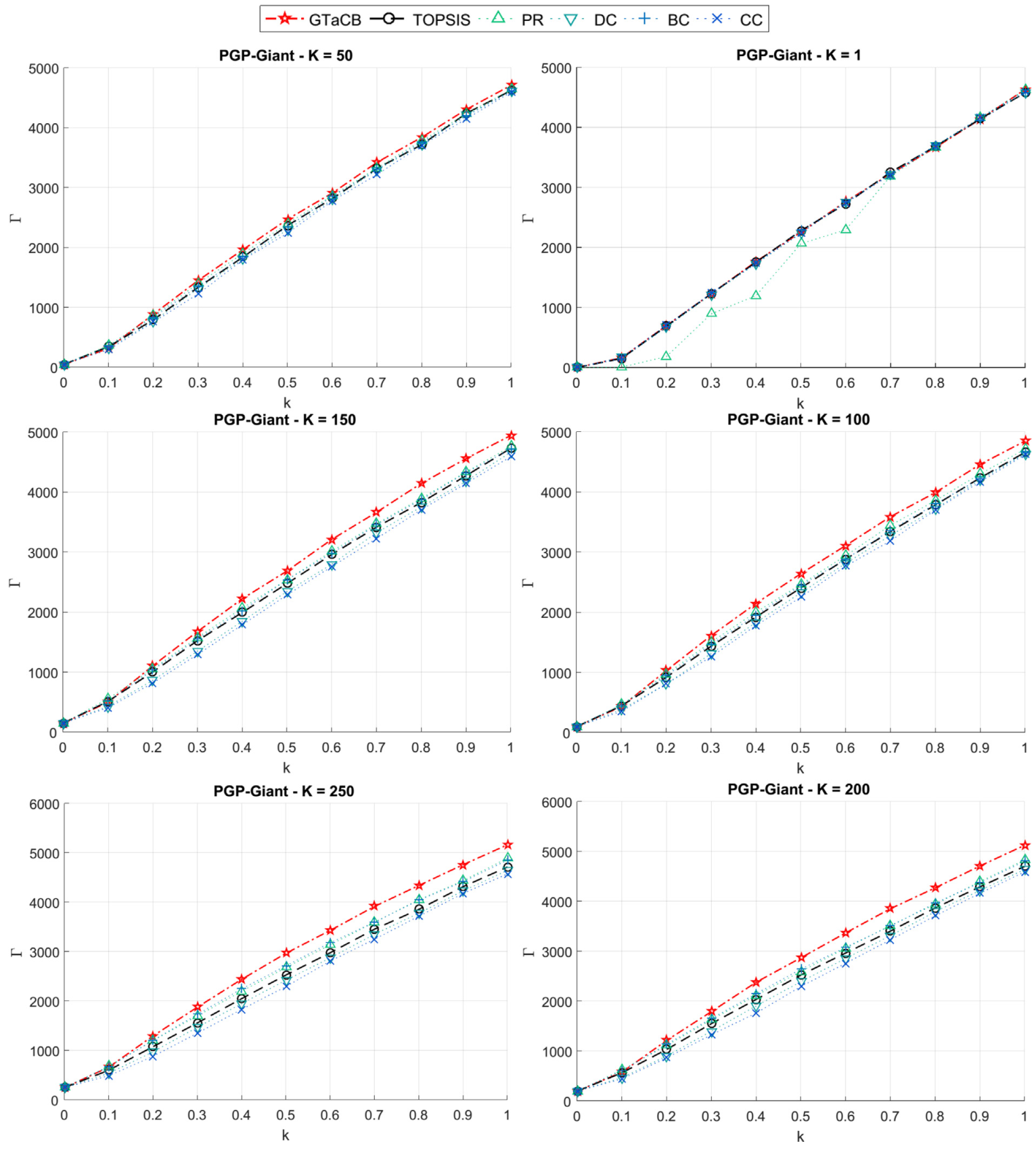}
    \caption{Infection result comparisons in PGP network}
    \label{fig:10}
\end{figure}

\begin{figure}
    \centering
    \includegraphics[width=.95\textwidth]{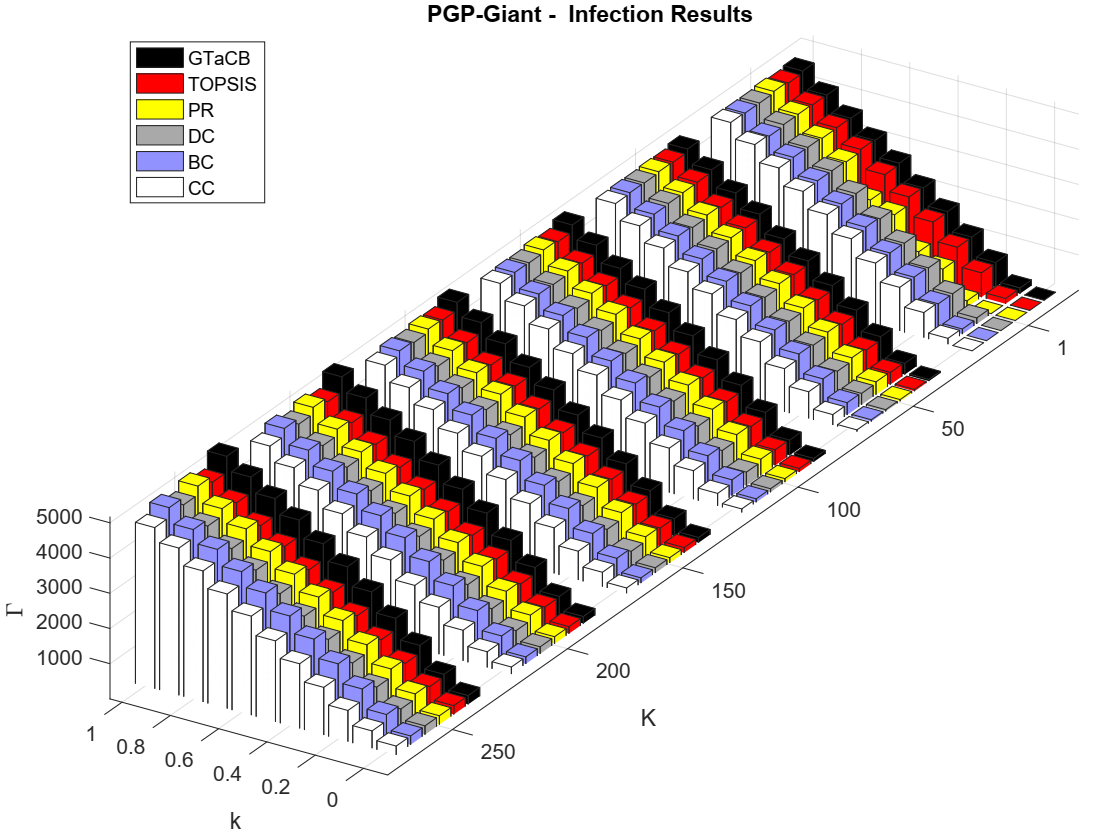}
    \caption{Infection result comparisons in 3D-bar chart of PGP network}
    \label{fig:11}
\end{figure}

In these comparisons, an algorithm whose set of seeds infects a larger number of nodes at the end of diffusion periods, has a higher quality. Thus, in \cref{fig:10}, where $K\geq50$, $\Gamma$ values of PGP network illustrates that GTaCB outperforms other algorithms when $K$ is greater than 0.1, in this condition, CC and DC’s seeds infect less than others and PR and BC follow GTaCB alongside each other. For instance, when $k=0.5$ and $K=200$, CC could infect near 2296 nodes at average, there were approximately 348 and 301 nodes less than those of BC (2644.88) and PR (2597.76), respectively, however GTaCB’s set of seeds infect more than 2871 nodes, puts this algorithm in the first place. By comparing \cref{fig:10} and \cref{fig:11}, it can clearly be seen that the gap between the algorithms’ qualities opens up, as either $K$ or $k$ increases. So, in PGP network as $K\geq 50$ and $k>0.1$, our algorithm significantly shows a higher performance in terms of diffusion quality. For instance, in our tests, when $K=250$ and $k= 0.4$, GTaCB’s seeds have infected 20.4\%, 10.4\%, 27.6\%, 8.3\% and 38.9\% more than those of TOPSIS, PR, DC, BC and CC, respectively.

\begin{figure}
    \centering
    \includegraphics[width=.95\textwidth]{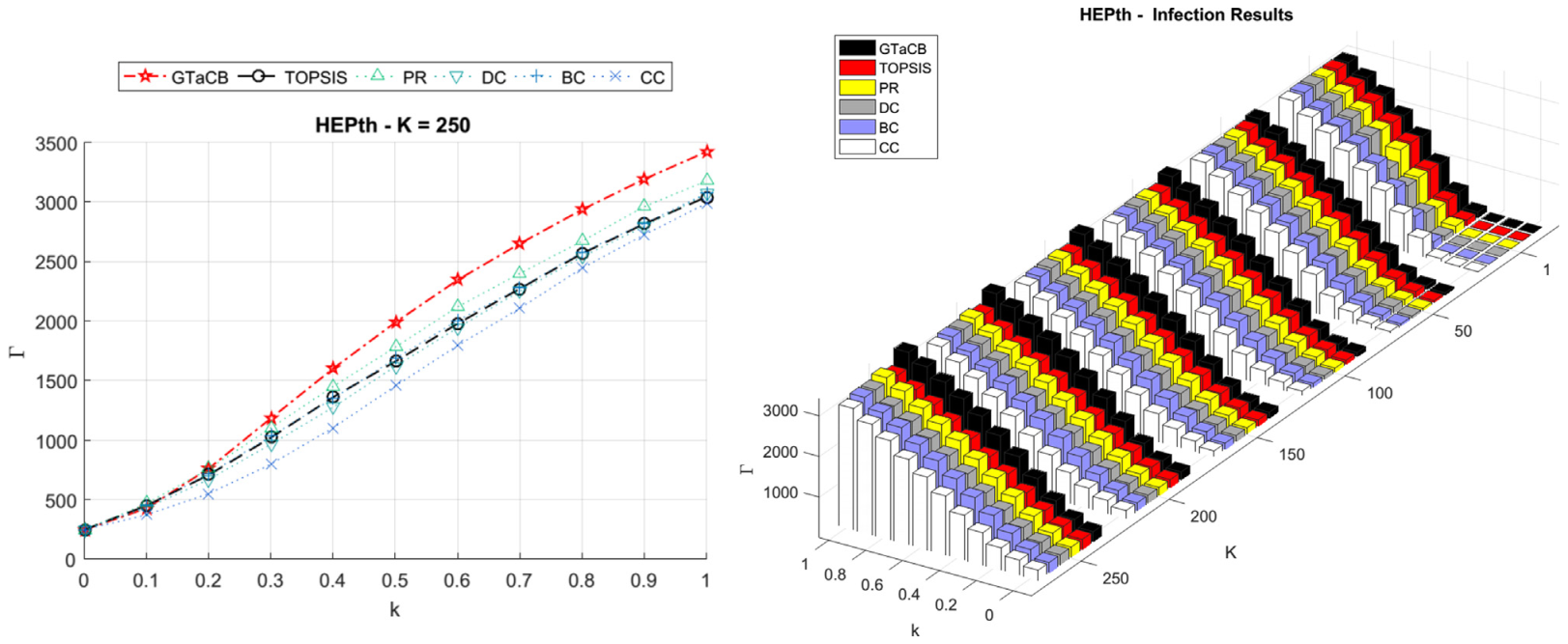}
    \caption{Infection result comparisons in HEP-th network}
    \label{fig:13}
\end{figure}

\begin{figure}
    \centering
    \includegraphics[width=.95\textwidth]{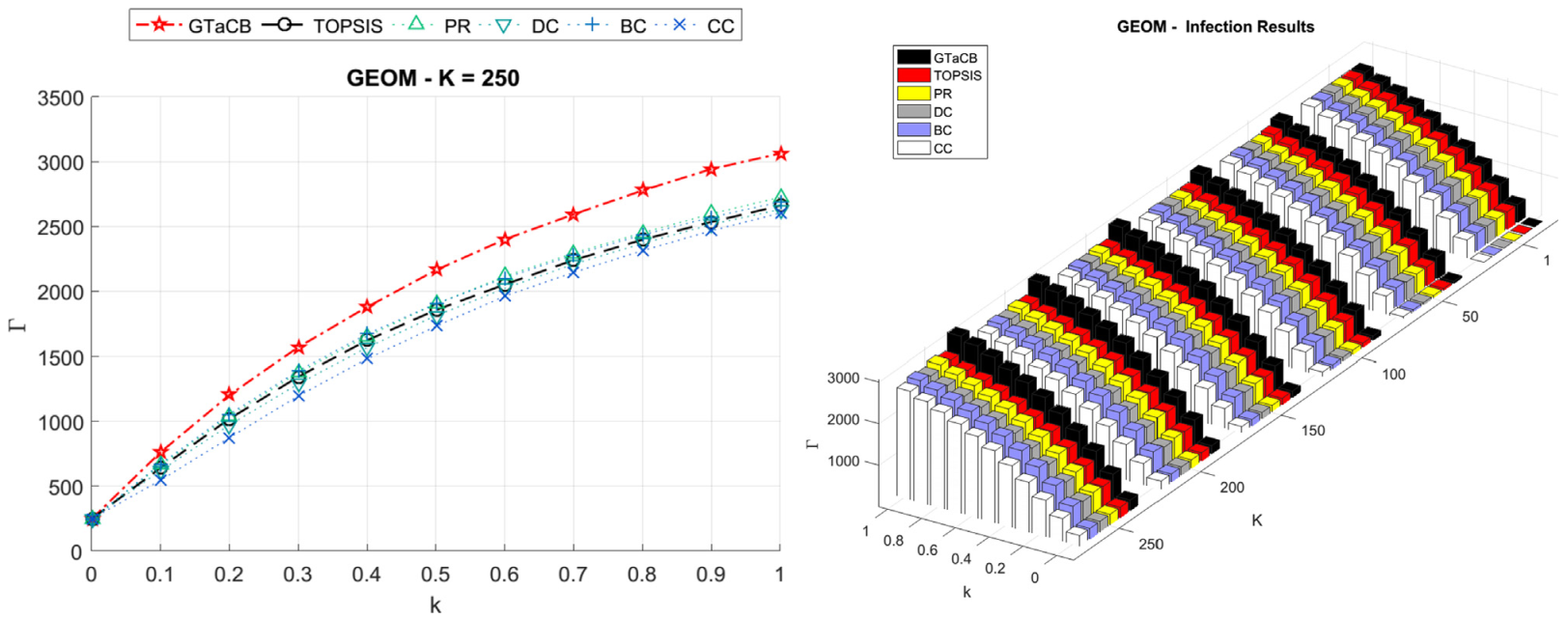}
    \caption{Infection result comparisons in GEOM network}
    \label{fig:14}
\end{figure}

\begin{figure}
    \centering
    \includegraphics[width=.95\textwidth]{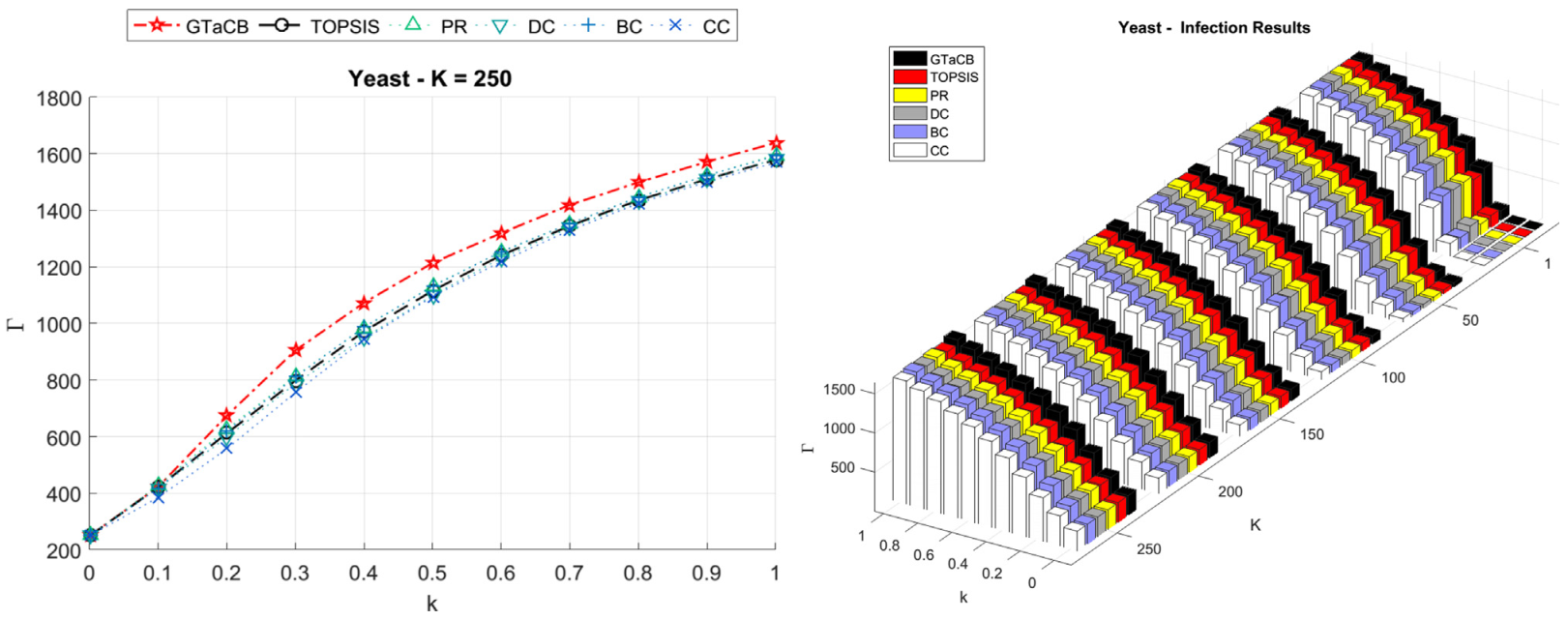}
    \caption{Infection result comparisons in YEAST network}
    \label{fig:15}
\end{figure}

\begin{figure}
    \centering
    \includegraphics[width=.95\textwidth]{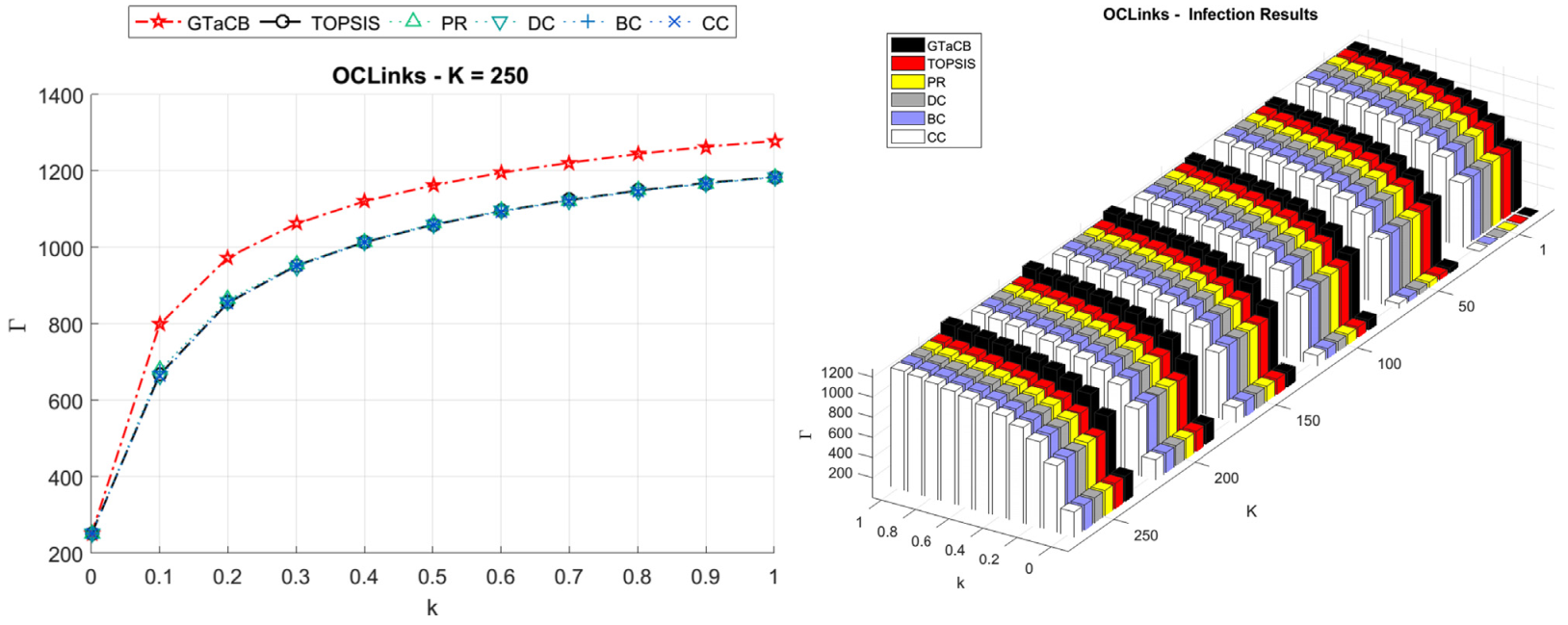}
    \caption{Infection result comparisons in OCLinks network}
    \label{fig:16}
\end{figure}

\begin{figure}
    \centering
    \includegraphics[width=.95\textwidth]{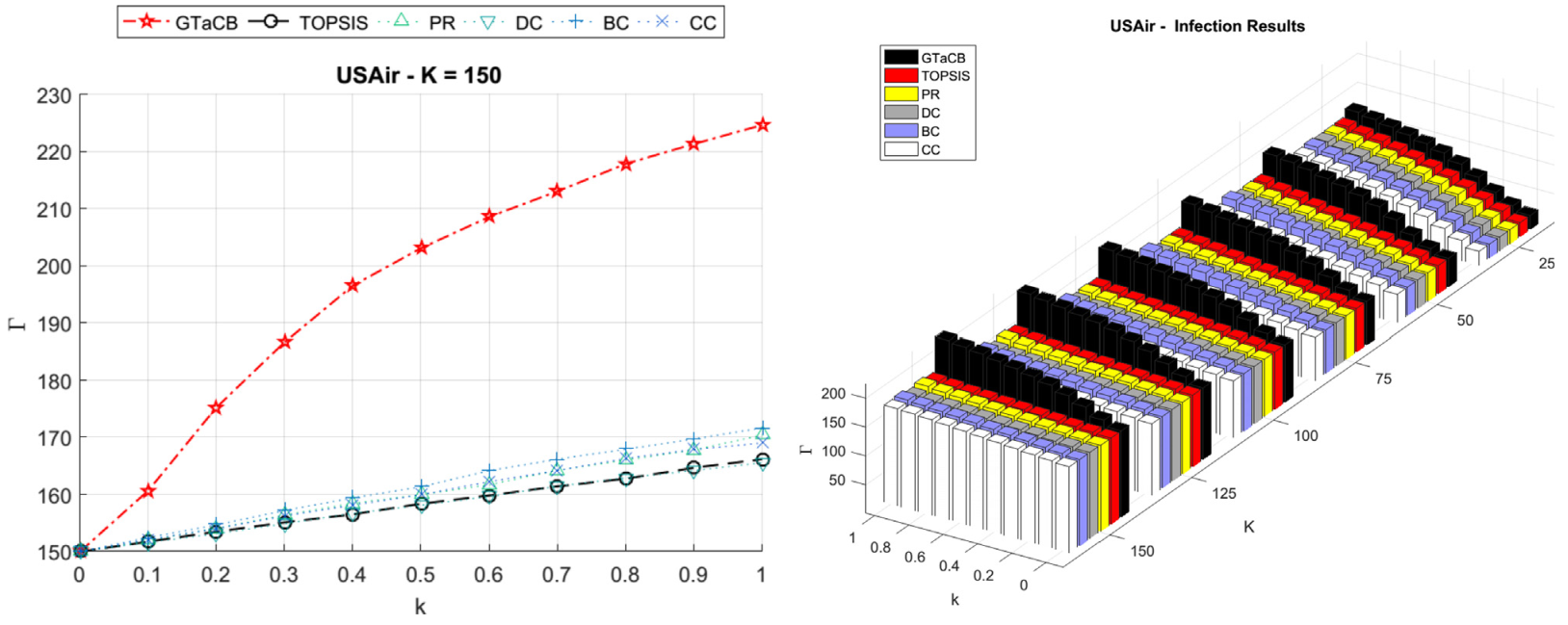}
    \caption{Infection result comparisons in USAir network}
    \label{fig:17}
\end{figure}

\begin{figure}
    \centering
    \includegraphics[width=.95\textwidth]{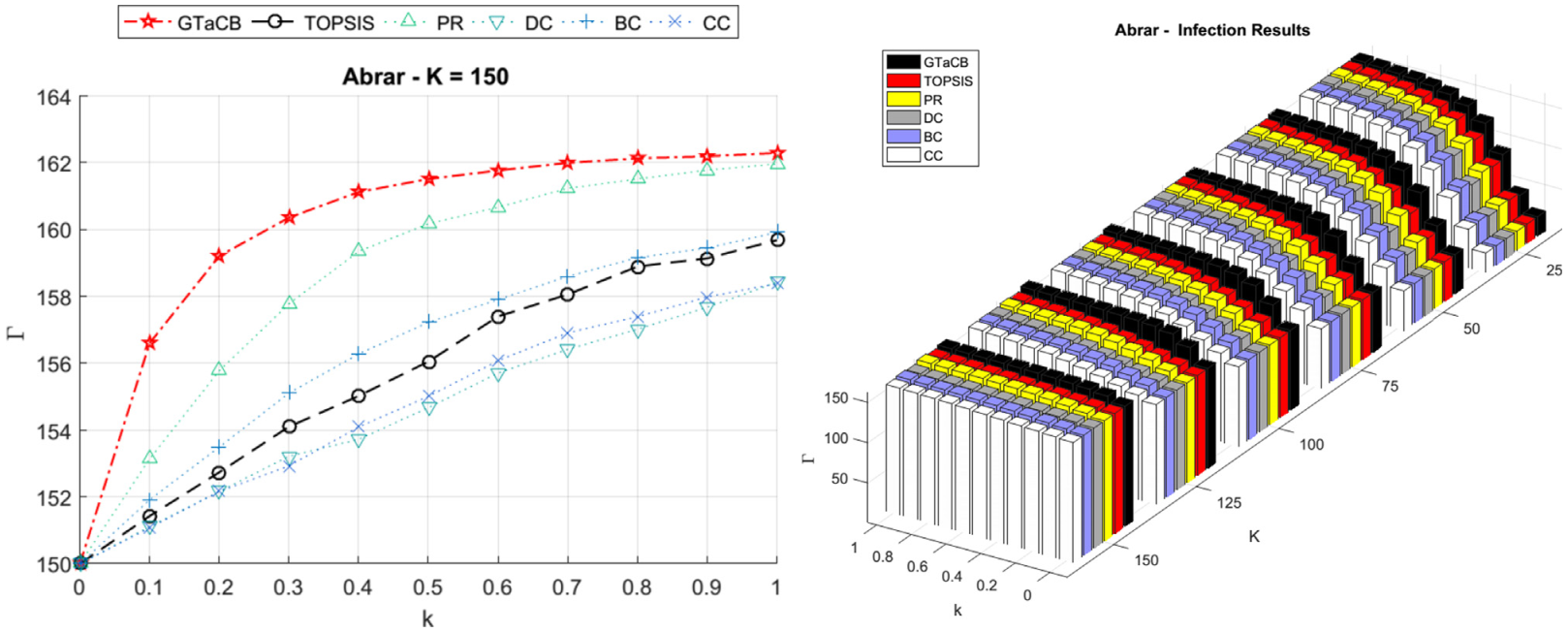}
    \caption{Infection result comparisons in Abrar network}
    \label{fig:18}
\end{figure}

Just alike PGP’s result, the epidemic spreading results in Slashdot, HEP-th, Geom, Yeast, OCLinks, USAir and Abrar networks exhibit the domination of GTaCB performance, chiefly as $K$ and $k$ increase; \cref{fig:12} to \cref{fig:18} illustrate the experimental results of these networks respectively:
\begin{figure}
    \centering
    \includegraphics[width=.95\textwidth]{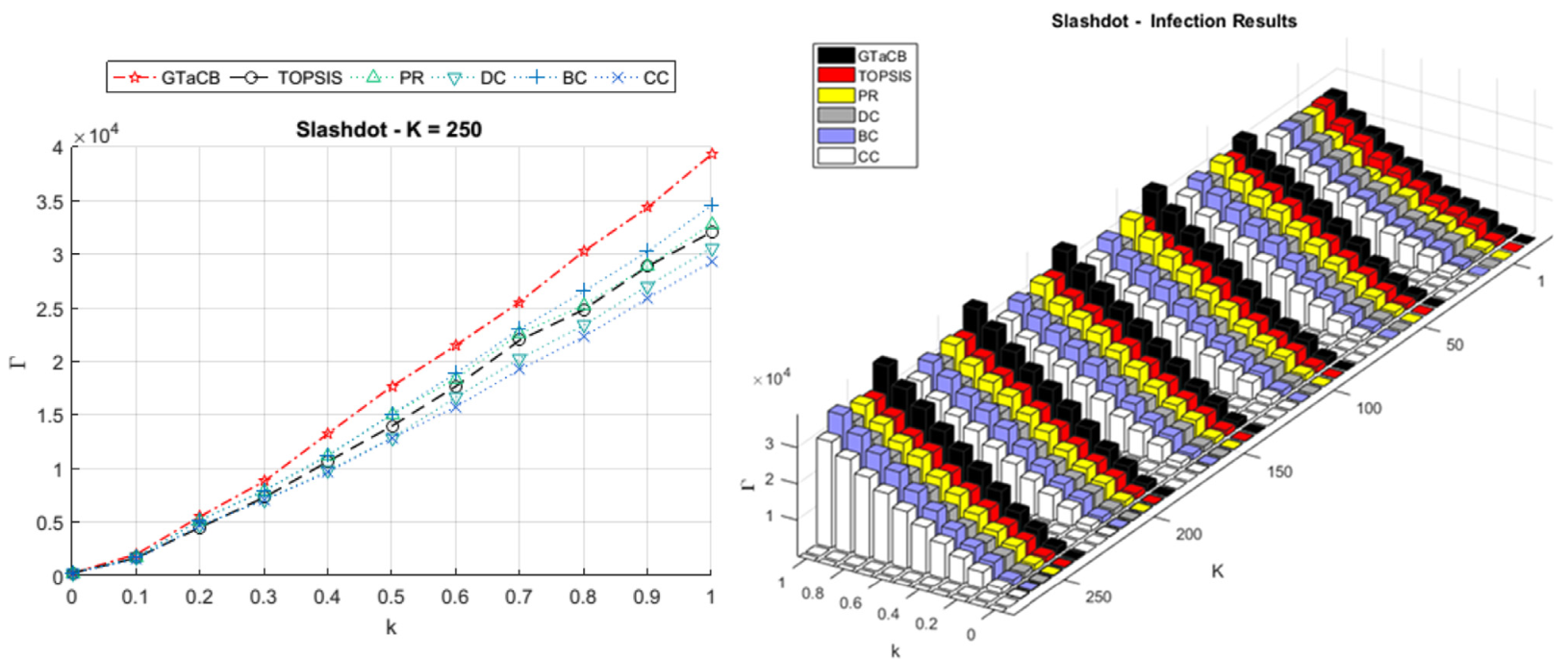}
    \caption{Infection result comparisons in Slashdot network}
    \label{fig:12}
\end{figure}

\begin{figure}
    \centering
    \includegraphics[width=.95\textwidth]{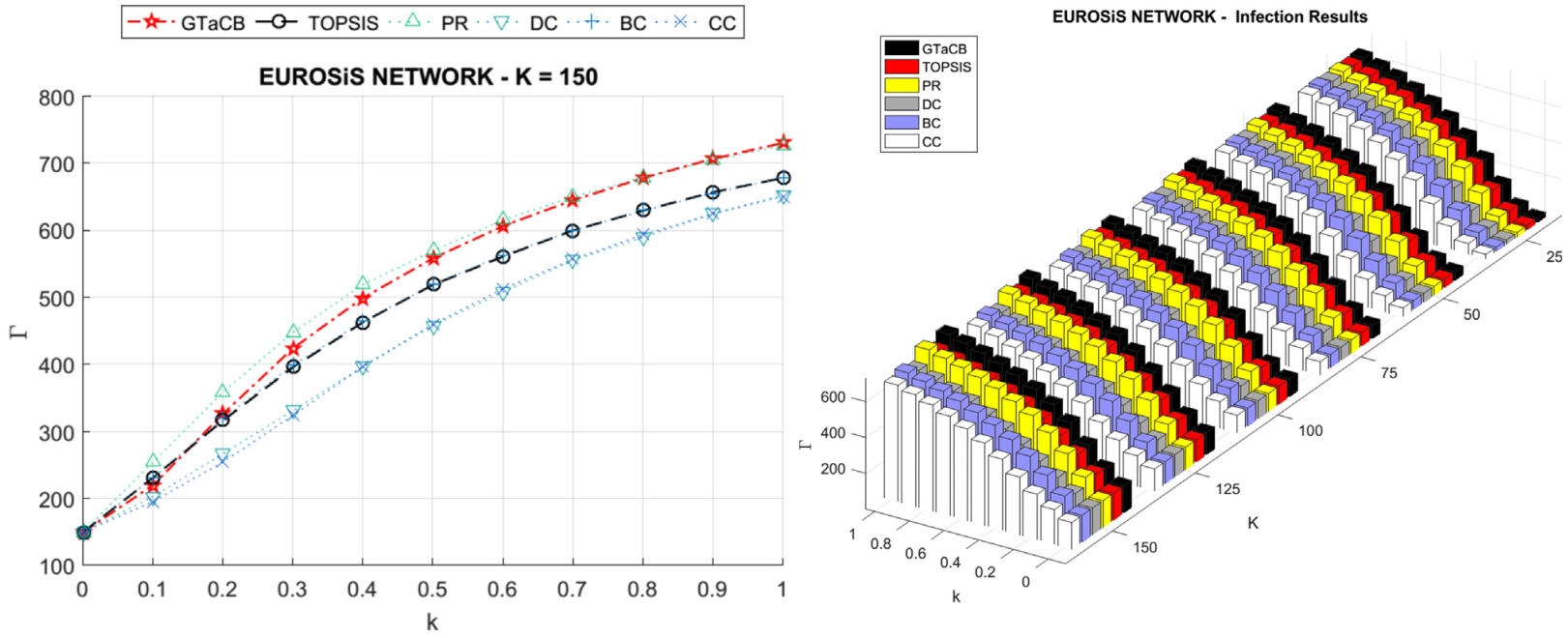}
    \caption{Infection result comparisons in EuroSiS network}
    \label{fig:19}
\end{figure}

Nevertheless, in our tests there was a network (EuroSiS) in which the quality of PR had dominion over its rivals. However, GTaCB was following it as the second highest quality in most of the situations can be seen in \cref{fig:19}.

By averaging $\Gamma_{i,K,k}$ values on $K$ and $k$ values, we summarized the results in \cref{tab:5}, to provide insights as to how GTaCB’s quality of diffusion outperforms the comparable algorithms through the examinations on Abrar, USAir, OCLinks, HEP-th, GEOM and PGP. Notwithstanding its quality, the proposed approach has a drawback of its runtime, due to the community detection techniques dullness and memory usage so that, the implementation of larger networks was almost impossible on the foresaid laptop, because of the “Out of Memory” MATLAB error.

\begin{table}[]
\label{tab:5}
\caption{Mean of infected nodes percentages on $K_r$ and $k_m$ conducted by different algorithms on the employed networks}
\begin{tabular}{lllllll}
\toprule
$\Gamma$ & GTaCB   & TOPSIS  & PR      & DC      & BC      & CC      \\
\midrule
Abrar                 & 88.22\% & 83.82\% & 85.09\% & 82.88\% & 84.37\% & 83.27\% \\
USAIR                 & 37.57\% & 31.02\% & 31.23\% & 30.89\% & 32.22\% & 30.95\% \\
EuroSiS               & 34.87\% & 33.80\% & 35.96\% & 31.43\% & 34.00\% & 31.16\% \\
OCLinks               & 52.28\% & 49.61\% & 49.65\% & 49.60\% & 49.63\% & 49.57\% \\
YEAST                 & 42.16\% & 40.72\% & 41.01\% & 40.62\% & 40.89\% & 40.22\% \\
Geom                  & 24.32\% & 22.09\% & 22.25\% & 21.77\% & 22.32\% & 21.40\% \\
HEP-th                & 19.37\% & 17.80\% & 18.32\% & 17.57\% & 17.90\% & 17.09\% \\
PGP                   & 24.01\% & 22.35\% & 22.59\% & 21.76\% & 22.77\% & 21.36\% \\
Slashdot              & 19.22\% & 16.70\% & 17.19\% & 15.74\% & 17.50\% & 15.44\%\\
\bottomrule
\end{tabular}
\end{table}

\subsection{Diffusion Speed of GTaCB}
\begin{figure}
    \centering
    \includegraphics[width=.95\textwidth]{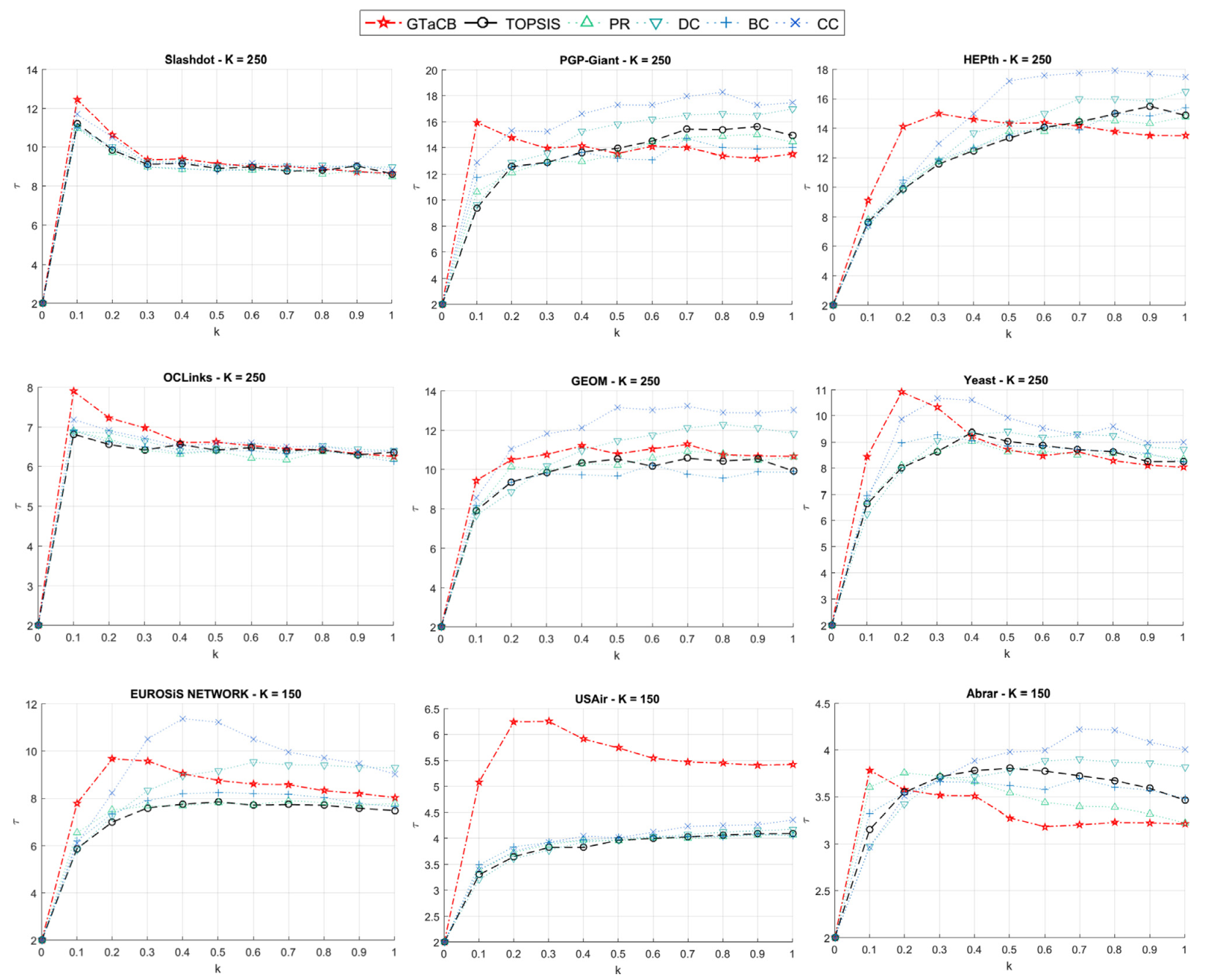}
    \caption{The comparison of $\tau$ values for the employed networks}
    \label{fig:20}
\end{figure}

As mentioned above, all $\tau$ values have been collected, and in \cref{fig:20} it is shown that as $K$ grows, $\tau$ values of GTaCB decrease as it did in PGP, HEP-th, Yeast and Abrar. It spreads in a shorter time in comparison with those of its rivals.

Its average $\tau$ value is almost more than others in many cases, as depicted in \cref{tab:6}. Since these values are depended on the diffusion quality, there is an ambiguity to judge which algorithms’ set of seeds infect more nodes in quickly. For example, in USAir network, diffusion process of DC’s set of seeds lasts 3.99 periods on average, which is the shortest time in comparison with those of others, but it infects only 30.89\% of the network. While, GTaCB’ set has the most average longest time of 5.91 periods, but it infects 37.57\% which outnumber others. Hence, we have defined another simple but applicable variable showing that how many nodes are infected through each period averagely. It makes us find out a kind of diffusion speed measure to evaluate the influential seed sets from a new point of view:
\begin{equation}
    \eta_{K,k}=\frac{\Gamma_{K,k}-K}{\tau_{K,k}}
\end{equation}

\begin{table}[]
\label{tab:6}
\caption{Mean of $\tau$ values on $K$ and $k$}
\begin{tabular}{lllllll}
\toprule
$\Bar{\tau}$      & GTaCB & TOPSIS & PR    & DC    & BC    & CC    \\
\midrule
Abrar    & 4.79  & 4.93   & 4.83  & 5.29  & 4.77  & 5.22  \\
USAIR    & 5.91  & 4.03   & 4.03  & 3.99  & 4.16  & 4.04  \\
EuroSiS  & 9.08  & 7.94   & 7.82  & 8.96  & 8.12  & 9.83  \\
OCLinks  & 6.97  & 6.61   & 6.58  & 6.73  & 6.58  & 6.73  \\
YEAST    & 10.12 & 9.70   & 9.68  & 9.85  & 9.72  & 10.18 \\
Geom     & 11.19 & 10.95  & 11.03 & 11.45 & 10.61 & 11.93 \\
HEP-th   & 15.26 & 14.19  & 14.30 & 14.75 & 14.62 & 15.98 \\
PGP      & 15.08 & 14.74  & 14.35 & 15.62 & 14.64 & 16.29 \\
Slashdot & 11.04 & 10.52  & 10.45 & 10.74 & 10.51 & 10.72\\
\bottomrule
\end{tabular}
\end{table}

Therefore, \cref{tab:7} compares the average diffusion speeds each algorithm performs on the employed networks. It shows how the sets that GTaCB identifies, outperformed the sets of others; however, again in EuroSiS our algorithm did not have better performance and this time, its value was only more than those of DC and CC, where PR peaked the diffusion speeds by far (approximately 44.5 nodes in each period).

\begin{table}[]
\label{tab:7}
\caption{Average $\eta$ values of compared algorithms showing that how many nodes are averagely infected within a period}
\begin{tabular}{lllllll}
\toprule
$\Bar{\eta}$       & GTaCB   & TOPSIS  & PR      & DC      & BC      & CC      \\
\midrule
Abrar    & 10.82   & 8.66    & 9.36    & 7.68    & 9.22    & 8.00    \\
USAIR    & 5.81    & 3.34    & 3.53    & 3.29    & 4.17    & 3.29    \\
EuroSiS  & 37.73   & 40.30   & 44.55   & 31.77   & 39.83   & 28.97   \\
OCLinks  & 118.66  & 117.26  & 118.04  & 115.38  & 117.98  & 115.07  \\
YEAST    & 85.68   & 82.80   & 83.76   & 80.87   & 83.84   & 77.31   \\
Geom     & 137.93  & 124.84  & 125.54  & 115.81  & 132.42  & 108.65  \\
HEP-th   & 88.87   & 82.55   & 85.77   & 77.08   & 82.26   & 69.32   \\
PGP      & 155.83  & 138.61  & 142.39  & 125.45  & 146.59  & 118.95  \\
Slashdot & 1480.15 & 1340.20 & 1393.50 & 1240.94 & 1411.35 & 1212.33\\
\bottomrule
\end{tabular}
\end{table}

\section{Conclusion}\label{sec:6}
Finding a set of influential nodes is of practical and theoretical importance in complex networks, specially to resolve a problem called “influence maximization”. In this paper, we proposed GTaCB, a new algorithm to find the set of initial nodes by distributing the network into $K$ sub-graphs via a community detection algorithm named, GCSC. And afterwards, by the implementation of a multi-attribute decision-making (MADM) technique known as, TOPSIS, to find the best node in each sub-graph with the aid of centrality measures as its attributes. The main novelty of the present paper is to cope with the influence maximization problem by utilizing four centrality measures of the nodes and considering the fact that the social networks are community based. To improve the previous studies in which the multi attribute techniques have been utilized, we used a community detection algorithm to separate the communities from each other to find the best influential nodes. 

In order to evaluate the performance of GTaCB in comparison to other algorithms, we have simulated diffusion of the chosen seeds, identified by each approach, through the employment of an SIR model. The experimental results show that in one hand, the set of nodes that GTaCB suggests, has a higher diffusion quality in 8 out of 9 networks. On the other hand, despite its long diffusion process, in most of the cases it is faster than others in terms of diffusion speed, in particular when infection rate is sufficiently high ($k\alpha>0.05$). Secondly, the results clearly illustrate that Jaccard Coefficient (JC) values between the sets identified by GTaCB and each of its rivals are considerably lower than JC values between the pair sets of the rest. It means that the majority of individual nodes, selected by the proposed algorithm, are not considered “influential” by the famous centrality measures. Apart from that, our experiments show that BC and PR compete rather effectively than DC and specially CC which was poorer in both diffusion quality and diffusion speed. 

From an application point of view, it means that the detection of the network communities and the selection of the seed nodes (using the four well-known centrality measures) in each community is a nice strategy to find the influential nodes. Additionally, it is suggested to develop a mathematical programming model for the problem and to compare the results of the GTaCB with optimal solution on small-sized datasets to show the optimal gap. 

\section*{Conflict of Interest}
The authors declare that they have no known competing financial interests or personal relationships that could have appeared to influence the work reported in this paper.

\section*{Acknowledgements}
None. No funding to declare.

\bibliography{references}

\end{document}